\documentclass[altaffilletter,aps,nofootinbib,twocolumn,prd,eqsecnum,preprintnumbers,superscriptaddress,10pt,floatfix]{revtex4-1}
\pdfoutput=1
\usepackage{subfigure}
\usepackage{graphicx}
\usepackage{amsmath}
\usepackage{amsfonts}
\usepackage{mathtools}
\usepackage{amssymb}
\usepackage{enumerate}
\usepackage{xcolor}
\usepackage{bm}
\usepackage{mathrsfs}
\usepackage{epstopdf}
\usepackage{url}
\usepackage{footnote}
\usepackage{textcomp}
\usepackage{dsfont}
\usepackage{ulem}
\usepackage{hyperref}
\usepackage{enumerate}   
\usepackage{appendix}
\usepackage{textcomp}
\usepackage{tipa}

\makeatletter
\newcommand*{\rom}[1]{\expandafter\@slowromancap\romannumeral #1@}
\makeatother

\begin{document}

\title{Resonant orbits of rotating black holes beyond circularity: discontinuity along parameter shift}

\author{Che-Yu Chen}
\email{b97202056@gmail.com}
\affiliation{RIKEN iTHEMS, Wako, Saitama 351-0198, Japan}
\affiliation{Institute of Physics, Academia Sinica, Taipei 11529, Taiwan}

\author{Hsu-Wen Chiang}
\email{b98202036@ntu.edu.tw}
\affiliation{Leung Center for Cosmology and Particle Astrophysics, National Taiwan University, Taipei 10617, Taiwan}

\author{Avani Patel}
\email{avani.physics@gmail.com}
\affiliation{Center of Astronomy and Gravitation, National Taiwan Normal University, Taipei 11677, Taiwan}
\affiliation{Department of Physics, National Taiwan Normal University, Taipei 11677, Taiwan}

\begin{abstract}
According to General Relativity, an isolated black hole in vacuum shall be described by the Kerr metric, whose geodesic equations are integrable. The violation of integrability leads to chaos for particles moving around the black hole. This chaotic dynamics could leave imprints on the associated gravitational waveform and could be tested with upcoming observations. In this paper, we investigate the chaotic orbital dynamics induced by the violation of a certain spacetime symmetry, the circularity. Specifically, we focus on the resonant orbits of a particular noncircular spacetime as an example and find that they form chains of Birkhoff islands on Poincar\'e surfaces of section. We compare the island structures with those generated in typical nonintegrable but circular spacetimes. The islands of stability induced by noncircularity appear asymmetric on the most common Poincar\'e surface of section at the equatorial plane. The asymmetric patterns of islands vary discontinuously when the spacetime parameters transit through integrable regions. The origin of such features is explained in the context of perturbation analysis by considering the orbits associated with stable fixed points on the section. Possible observational implications about testing circularity through gravitational wave detection are discussed.           

\end{abstract}

\maketitle

\section{Introduction}

The direct detection of gravitational waves emitted by binary merger events \cite{LIGOScientific:2016aoc} and the images of supermassive black holes released by the Event Horizon Telescope \cite{EventHorizonTelescope:2019dse,EventHorizonTelescope:2022wkp} are of epoch-making significance in the field of black hole physics. With future advancements in detectors, it becomes possible to probe what is happening in the vicinity of black holes. In particular, these observations may help to reveal whether the black holes follow General Relativity (GR) in our universe. More explicitly, we can directly test the Kerr hypothesis, i.e., to what extent may the Kerr geometry describes these extremely compact astrophysical objects.

According to GR, the exterior spacetime of an isolated and spinning black hole should be the Kerr geometry. On top of the apparent symmetries of stationarity and axisymmetry, the Kerr geometry possesses additional symmetry properties. First, the Kerr spacetime has a well-defined equatorial plane with respect to which the spacetime is symmetric. Second, the Kerr spacetime has a hidden symmetry that allows for the separability of geodesic equations. The separability of geodesic equations indicates the existence of an additional constant of motion other than the ones that correspond to stationarity and axisymmetry, called the Carter constant \cite{Carter:1968rr}. For orbital dynamics on pseudo-Riemannian manifolds, the separability of geodesic equations implies the integrability in the Liouville sense{\footnote{The separability here means that the radial and latitudinal components of the geodesic equation can be separated. For the Kerr metric, the wave equations also turn out to be separable. When considering metrics beyond Kerr, the separability of geodesic equations does not imply the separability of wave equations \cite{Chen:2019jbs,Papadopoulos:2020kxu}.}}. Therefore, the orbital dynamics has no chaos. The violation of the Kerr hypothesis, which could be due to, for example, the astrophysical environments, the existence of companions, physics beyond GR, or putative quantum gravitational corrections, may break the aforementioned symmetry properties. Therefore, the Kerr hypothesis could be tackled by directly attacking the individual symmetry of the Kerr spacetime, i.e., equatorial reflection symmetry, Liouville integrability, etc. The violation of any of these symmetries directly implies the violation of the Kerr hypothesis. 

One specific possibility in this direction is to test the Liouville integrability via gravitational waves. Specifically, the nonintegrability of spacetimes may induce chaos in orbital dynamics. The existence of chaos in the dynamics of a spinning test particle in Schwarzschild background and its imprints on gravitational waveforms were primarily studied in \cite{Suzuki:1996gm,Suzuki:1999si}, while the possibility of chaotic geodesics in a non-Kerr spacetime was studied in \cite{Gair:2007kr}. Recently, in Refs.~\cite{Destounis:2021rko,Destounis:2021mqv,Apostolatos:2009vu,Lukes-Gerakopoulos:2010ipp}, it has been shown that chaotic orbital dynamics may leave a particular imprint on the gravitational waves emitted by the extreme-mass-ratio inspirals (EMRIs), which consist of a stellar size object gradually spiraling toward a supermassive black hole. The extreme mass ratio of EMRI systems implies that, at the leading order of the mass ratio, the trajectory of the stellar object can be approximated by the geodesic equations defined in the spacetime of the supermassive black hole. If the supermassive black hole violates the Kerr hypothesis by breaking the Liouville integrability, the chaotic features would directly appear in the geodesic dynamics and could be identified through gravitational waves. In fact, the gravitational waves emitted by EMRIs are one of the main targets of future space-based gravitational wave detectors \cite{Gair:2017ynp}, such as LISA \cite{Glampedakis:2005hs,LISA:2017,LISA:2022yao,LISA:2022kgy}. Therefore, it has become timely and crucial to investigate the possibility of testing the Liouville integrability through gravitational waves.

Having mentioned the possibility of probing Liouville integrability through gravitational waves, one naturally asks if we may extract more information about the geometry through the manifested chaos. Along this line of thought, a geometrical property called circularity stands out. Suppose we have a stationary and axisymmetric spacetime with two Killing vectors. Their associated 1-forms are denoted by $k$ and $\eta$. The spacetime metric is circular if there exist two-surfaces that are everywhere orthogonal to the surfaces of transitivity, i.e., the surfaces spanned by the Killing vectors. In GR, this geometric property induces some constraints that the energy-momentum tensor must satisfy. In the case of fluid, these constraints imply that there is no convective motion, and only the circular motion around the axis of symmetry is allowed \cite{Gourgoulhon:1993,Ioka:2003dd,Ioka:2003nh,Gourgoulhon:2010ju,Birkl:2010hc}. According to Frobenius's theorem, the circularity condition is identical to the following integrability conditions \cite{Ayon-Beato:2005mje}:
\begin{align}
\mathcal{C}_1\equiv k\wedge\eta\wedge dk=0\,,\nonumber\\
\mathcal{C}_2\equiv k\wedge\eta\wedge d\eta=0\,.\label{circularcri}
\end{align}
Violating any of these two criteria would imply that the spacetime is noncircular. 

The motivation for noncircularity was originally to incorporate the convective flow of matters, such as the meridional circulation on neutron stars with toroidal magnetic fields, into GR \cite{Gourgoulhon:1993}. However, if any signatures of noncircularity appear in an isolated black hole system, which is supposed to be vacuum, they likely originate from non-GR effects \cite{VanAelst:2019kku,Fernandes:2023vux}. In Refs.~\cite{BenAchour:2020fgy,Anson:2020trg,Anson:2021yli}, a class of noncircular black hole spacetimes was constructed by disformal transformations in the context of scalar-tensor theories of gravity. In fact, noncircularity seems generic among metrics generated by such a solution-generating method \cite{Minamitsuji:2020jvf}. On the other hand, it was proved in Ref.~\cite{Xie:2021bur} that, under some assumptions, the black hole solutions obtained in generic effective theories of gravity have to satisfy circularity conditions if they can be attained perturbatively from a given circular solution in the GR limit. This is consistent with Ref.~\cite{Nakashi:2020phm}, in which explicit examples were provided. The result of Ref.~\cite{Xie:2021bur} places a strict theoretical constraint on how noncircularity may occur in theories beyond GR. However, it does not exclude the possibility that noncircular spacetimes may arise in scenarios where the assumptions made in Ref.~\cite{Xie:2021bur} are not satisfied. Also, noncircular spacetimes may appear in the branches of solutions that can not perturbatively reduce to the GR limit.

In addition to considering noncircular solutions obtained in individual theories of gravity, one can adopt a more phenomenological perspective and construct models of noncircular spacetimes in a theory-agnostic manner. In Refs.~\cite{Eichhorn:2021etc,Eichhorn:2021iwq}, a class of phenomenological metrics for regular black hole spacetimes was constructed. The model is characterized by an additional non-GR parameter that not only regularizes the spacetime but also makes the spacetime noncircular. Rotating black hole spacetimes in the framework of asymptotically safe gravity may have similar properties \cite{Held:2019xde,Eichhorn:2022bgu}\footnote{For detailed review on the black holes in asymptotically safe gravity please c.f. \cite{Platania:2023srt}.}. Recently, a general parameterized metric for noncircular spacetimes has been proposed \cite{Delaporte:2022acp}. These parameterized and phenomenological frameworks allow for the direct explorations of special observational features generated by noncircularity. It has been recently demonstrated that the images cast by a noncircular black hole may have some novel features, such as fractal structures \cite{Long:2020wqj}, tiny cusp-like structures on the shadow boundary, and reflection asymmetry of shadow boundaries on images that are not face-on \cite{Held:2019xde,Eichhorn:2022bgu,Delaporte:2022acp,Zhang:2023bzv}. The possibility of testing the noncircularity of the black hole spacetime of Ref.~\cite{BenAchour:2020fgy,Anson:2020trg} using orbiting pulsars was proposed in Ref.~\cite{Takamori:2021atp}.

In this work, as what has been done for testing Liouville integrability through gravitational waves, we will focus on the orbital dynamics of a massive particle around noncircular black holes. In general, noncircularity would break the Liouville integrability, and hence, chaos would appear. It has been shown in Ref.~\cite{Zhou:2021cef} to be so in the particular noncircular spacetime of Refs.~\cite{BenAchour:2020fgy,Anson:2020trg}. In this paper, by examining the structures of Birkhoff islands on Poincar\'e surfaces of section of orbital dynamics, we will show that the chaotic features generated by noncircularity differ substantially from those of circular spacetimes, especially for resonant orbits. We will show that the region in phase space where chaos emerges, e.g., the location of islands, depends sensitively on whether the spacetime is circular rather than the strength of noncircularity. As a result, even if the amount of noncircularity is tiny, it would already generate considerable effects on the resonant orbits. This would open a new window for testing circularity through the gravitational waves emitted by EMRIs.

The rest of this paper is organized as follows. In sec.~\ref{NC-spacetime}, we will first introduce a phenomenological model of noncircular black hole spacetimes that serves as the basis of our analysis. Then, we will briefly review the Poincar\'e surface of section and Birkhoff islands, which will be widely used in this paper to investigate chaotic orbital dynamics. A quick comment on the relation between (non)circularity and (non)integrability will be made at the end of this section. In sec.~\ref{sec.numerical}, we will demonstrate our main numerical results. We will show how noncircularity induces novel chaotic features on Poincar\'e surfaces of section, which do not appear in nonintegrable but circular cases. Then, in sec.~\ref{sec.pertur}, we will adopt a perturbation analysis to provide a quantitative description of these novel features. The overall interpretations of our numerical results and potential implications of gravitational wave observations will be discussed in sec.~\ref{sec.implication}. Finally, we will conclude in sec.~\ref{sec.conclusion}.   

\section{Orbital dynamics in noncircular spacetimes}\label{NC-spacetime}
In this section, we will review the ingredients needed to investigate orbital dynamics in noncircular spacetimes. We will first briefly introduce a class of non-Kerr spacetimes that will be the main focus of this work. This class of spacetimes contains new physics beyond GR that breaks circularity manifested by the Kerr spacetime. Then, we will focus on the geodesic dynamics of a massive particle moving in the non-Kerr spacetime and introduce the technical tools for identifying the chaotic features of the orbits. The relation between noncircularity and chaotic orbital dynamics will be elucidated as well.  

\subsection{Noncircular metric beyond Kerr}

In Ref.~\cite{Eichhorn:2021etc,Eichhorn:2021iwq,Delaporte:2022acp}, a class of phenomenological models of noncircular black hole spacetimes has been constructed. The procedure starts with the Kerr metric written in the Boyer–Lindquist coordinates $(t,r,\chi,\varphi_{\textrm{BL}})$ with $\chi=\cos{\theta}$:
\begin{align}
ds_{\textrm{K}}^2=&-\left[1-\frac{2Mr}{\rho(r,\chi)^2}\right]dt^2-\frac{4Mar}{\rho(r,\chi)^2}\left(1-\chi^2\right)dtd\varphi_\textrm{BL}\nonumber\\
&+\frac{\rho(r,\chi)^2}{\Delta_\textrm{K}(r)}dr^2+\frac{\rho(r,\chi)^2}{1-\chi^2}d\chi^2\nonumber\\
&+\frac{1-\chi^2}{\rho(r,\chi)^2}\left[\left(r^2+a^2\right)^2-a^2\Delta_\textrm{K}(r)\left(1-\chi^2\right)\right]d\varphi_\textrm{BL}^2\nonumber
\end{align}
where $\rho(r,\chi)^2\equiv r^2+a^2\chi^2$ and $\Delta_\textrm{K}(r)\equiv r^2-2Mr+a^2$, with black hole mass $M$ and spin $a$. The above Kerr line element can be rewritten in ingoing Kerr coordinates $(u,r,\chi,\varphi)$ through the following coordinate transformations on $(t,\varphi_\textrm{BL})$ that leave $(r,\chi)$ coordinates and $(\partial_t , \partial_\varphi)$ basis intact:
\begin{equation}
dt=du-\frac{r^2+a^2}{\Delta_\textrm{K}(r)}dr\,,\quad
d\varphi_\textrm{BL}=d\varphi-\frac{a}{\Delta_\textrm{K}(r)}dr\,.
\end{equation}
In this coordinate system, one then promotes the constant mass $M$ to a mass function $M(r,\chi)$ that depends both on $r$ and $\chi$ \cite{Eichhorn:2021etc,Eichhorn:2021iwq,Delaporte:2022acp}. The new spacetime metric so obtained can be expressed as
\begin{align}
ds^2=&-\left[1-\frac{2M(r,\chi)r}{\rho(r,\chi)^2}\right]du^2+2dudr+\frac{\rho(r,\chi)^2}{1-\chi^2}d\chi^2\nonumber\\
&-\frac{4M(r,\chi)ar}{\rho(r,\chi)^2}\left(1-\chi^2\right)dud\varphi-2a\left(1-\chi^2\right)drd\varphi\nonumber\\
&+\frac{1-\chi^2}{\rho(r,\chi)^2}\left[\left(r^2+a^2\right)^2-a^2\Delta(r,\chi)\left(1-\chi^2\right)\right]d\varphi^2\,,
\label{metric}
\end{align}
where $\Delta(r,\chi)\equiv r^2-2M(r,\chi)r+a^2$ is promoted as well. The new metric still respects stationarity and axisymmetry as for Kerr one because Eq.~\eqref{metric} does not have explicit $u$ and $\varphi$ dependence. However, it describes a class of non-Kerr spacetimes \cite{Eichhorn:2021etc,Eichhorn:2021iwq,Delaporte:2022acp} that is not circular anymore. In particular, such a metric differs fundamentally from the class of non-Kerr spacetimes obtained by promoting $M$ to $M(r,\chi)$ directly in the Boyer–Lindquist coordinates because the latter is always circular.

Extending from the metric of Refs.~\cite{Eichhorn:2021etc,Eichhorn:2021iwq,Delaporte:2022acp}, in this work we consider the following mass function
\begin{equation}
M(r,\chi)\equiv \frac{M}{1+48M^2l_{\textrm{NP}}^4/\left(r^2+\zeta a^2\chi^2\right)^3}\,,  \label{eq:mass_func}
\end{equation}
which reduces to the one of Refs.~\cite{Eichhorn:2021etc,Eichhorn:2021iwq,Delaporte:2022acp} when the dimensionless parameter $\zeta$ goes to unity. The parameter $l_\textrm{NP}$ represents the length scale under which the spacetime gets modified by some new physics beyond GR \cite{Eichhorn:2021iwq}. In this work, we allow $\zeta$ to vary within $[-1,1]$. As will be shown later, both $l_\textrm{NP}$ and $\zeta$ directly control the onset of noncircularity. The addition of the parameter $\zeta$ allows us to investigate the chaotic behavior of geodesics when the metric transits between circularity and noncircularity.

The non-Kerr spacetime of Eq.~\eqref{metric} has some important features. First, the singularity that would appear at $r=\chi=0$ in the Kerr spacetime is resolved\footnote{There could be a singular surface $48 l_\textrm{NP}^4 + \left(r^2 + a^2 \zeta \chi^2\right)^3=0$ when $\zeta<0$. But it does not affect the validity of our analysis and results.}. Second, the geodesics of massive objects are Liouville integrable in the following three scenarios:
\begin{enumerate}
    \item When $a=0$, i.e., when the spacetime becomes spherically symmetric.
    \item When $l_\textrm{NP}=0$, i.e., when the spacetime reduces back to Kerr one.
    \item When $\zeta=0$, i.e., when the mass function depends only on $r$.\footnote{According to Ref.~\cite{Eichhorn:2021iwq}, in this scenario the spacetime still receives corrections that do not satisfy locality. The correction can be interpreted as an isotropic screening to the bare mass.}
\end{enumerate}
Third, because in general, the mass function depends on both $r$ and $\chi$, the spacetime is no longer circular. With $k\equiv g_{u\mu}dx^\mu$ and $\eta\equiv g_{\varphi\mu}dx^\mu$, the quantities inside the circularity criteria of Eq.~\eqref{circularcri} are evaluated explicitly as
\begin{align}
\mathcal{C}_1\propto\frac{a^2\zeta l_\textrm{NP}^4 r\chi\left(r^2+a^2\zeta\chi^2\right)^2\left(1-\chi^2\right)}{\left(48 M^2 l_\textrm{NP}^4+\left(r^2+a^2\zeta\chi^2\right)^3\right)^2}\,,\nonumber\\
\mathcal{C}_2\propto\frac{a^3\zeta l_\textrm{NP}^4 r\chi\left(r^2+a^2\zeta\chi^2\right)^2\left(1-\chi^2\right)}{\left(48 M^2 l_\textrm{NP}^4+\left(r^2+a^2\zeta\chi^2\right)^3\right)^2}\,,
\end{align}
and are non-vanishing unless certain conditions are satisfied. In particular, the spacetime meets the circularity criteria in those scenarios where geodesics are integrable.

As will be mentioned later in sec.~\ref{sec:noncircularity_chaos}, noncircular spacetimes do not have any hidden nontrivial Killing tensor{\footnote{The trivial one is the metric tensor $g_{\mu\nu}$ itself.}}, and chaos may appear in the orbital dynamics. In the corollary, the Liouville integrability implies circularity, as demonstrated in the above analysis. In fact, the violation of circularity not only leads to chaotic orbital dynamics but also to novel orbital characteristics unheard-of before in circular spacetimes. Later, we will explicitly show this both numerically and analytically.

\subsection{Orbital dynamics: Poincar\'e surface of section and Birkhoff islands}

Before we move further to a detailed analysis of (chaotic) orbital dynamics, let us briefly review the relevant theoretical basis for such systems.

Let us consider the orbital dynamics of a massive test particle of mass $\mu$ moving in a stationary and axisymmetric spacetime described by a metric $g_{\mu\nu}$. For simplicity, we will set $\mu=1$ in the rest of the paper\footnote{The conjugate momenta shall be scaled accordingly for observational purposes.}. The orbital Lagrangian of the test particle is
\begin{equation}
\mathcal{L}=\frac{1}{2}g_{\mu\nu}\dot{x}^\mu\dot{x}^\nu\,,
\end{equation}
where the overhead dot denotes the derivative with respect to the proper time $\tau$. Through the Lagrangian, one can construct the conjugate momenta $p_\mu$. The stationarity and axisymmetry of the spacetime manifest themselves through the metric being independent of time and azimuthal angle ($u$ and $\varphi$ in the case of Eq.~\eqref{metric}), hence implying the existence of two Killing vectors. These Killing vectors correspond to two constants of motion: the energy $E$ and the azimuthal angular momentum $L_z$ of the particle measured at spatial infinity. With these two constants of motion, the equations of motion that govern the orbital dynamics are in general two coupled second-order differential equations involving two remaining coordinates ($r$, $\chi$ in the case of Eq.~\eqref{metric}) and their derivatives, subject to a Hamiltonian constraint. This Hamiltonian constraint is associated with the conservation of the rest mass of the test particle. Here, we mainly focus on the evolution of bound orbits around the black hole, which can be obtained by solving the equations of motion with appropriate choices of $E$, $L_z$, and initial conditions.

One of our central goals is to demonstrate the chaotic nature of the orbital dynamics of a noncircular spacetime. It turns out that this can be more easily achieved by constructing the Poincar\'e surface of section. Essentially, with the Hamiltonian constraint, the trajectory of a bound orbit around black holes lies in a three-dimensional phase space. A Poincar\'e surface of section is essentially a two-dimensional surface embedded in this three-dimensional phase space. In the literature regarding the demonstration of chaotic orbital dynamics, the Poincar\'e surface of section is usually chosen as the $(r,\dot{r})$ or $(r,p_r)$ surfaces on the equatorial plane $\chi=0$ \cite{Takahashi:2008zh,Apostolatos:2009vu,Lukes-Gerakopoulos:2010ipp,Contopoulos:2011dz,Lukes-Gerakopoulos:2012qpc,Brink:2015roa,Zelenka:2017aqn,Lukes-Gerakopoulos:2017jub,Cardenas-Avendano:2018ocb,Lukes-Gerakopoulos:2021ybx,Chen:2022znf,Mukherjee:2022dju,Destounis:2023gpw,Destounis:2020kss,Deich:2022vna,Destounis:2023khj,Destounis:2023cim}.

\begin{figure}
    \centering
    \includegraphics[scale=0.08]{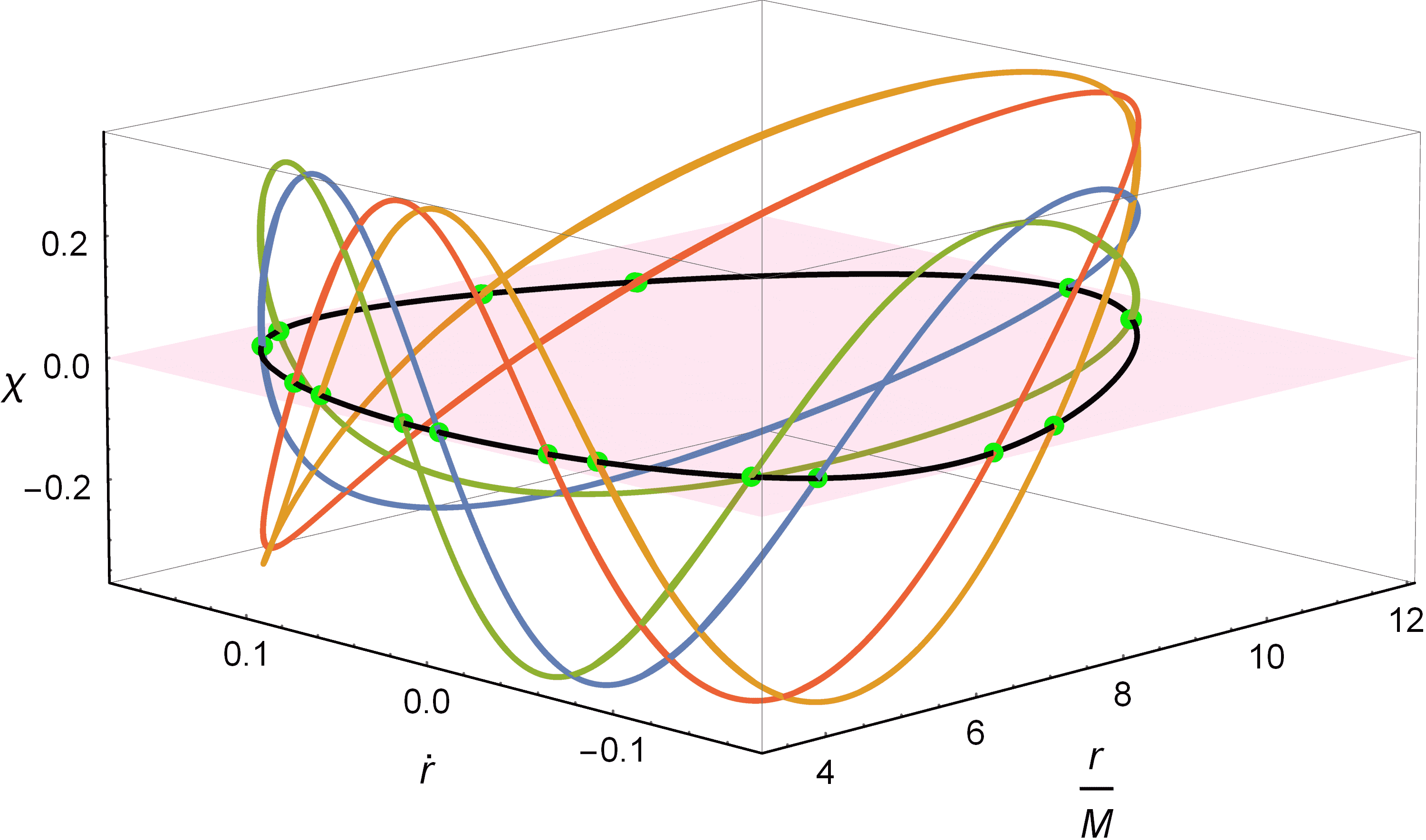}
    \caption{The phase space trajectories of the $1/2$-resonant orbits with $E=0.942$ and $L_z=2.76M$ for the integrable subset of the metric of Eq.~\eqref{metric}. We choose specifically $l_\textrm{NP}=0.4M$, $\zeta = 0$, and $a=0.66M$. The black curve is the closed curve formed by all $1/2$-resonant orbits on the Poincar\'e section at the equatorial plane (pink). The colored curves are some samples of orbits with different initial conditions. The green points represent the registered points of these orbits on the Poincar\'e surface of section.}
    \label{fig:PS3D_zeta0}
\end{figure}

Suppose we choose the $(r,\dot{r})$ surface on the equatorial plane as the Poincar\'e surface of section. With the time evolution of a bound trajectory, the orbit would keep intersecting the equatorial plane. We then record the points when the orbit pierces the plane from the south, i.e., $\dot{\chi}>0$. For a spacetime that possesses a hidden symmetry corresponding to a nontrivial rank-2 Killing tensor, the orbital dynamics is Liouville integrable. The hidden symmetry leads to one additional constant of motion, the Carter constant, that further confines the orbital trajectory in the phase space. For an integrable orbital system, if the orbit oscillates along both the radial and latitudinal directions, with a rational oscillation frequency ratio between the two, the orbit would intersect the equatorial plane at finite locations, registered as some points on the Poincar\'e surface of section. The number of points is determined by the frequency ratio. These orbits are called resonant orbits. On the other hand, if the frequency ratio is irrational, the orbits are called quasiperiodic and they form invariant tori in the phase space. Each of such orbits leaves a closed curve, called invariant curves, on the Poincar\'e surface of section. Typical invariant curves in an integrable system, e.g., the orbital dynamics of the Kerr spacetimes, can be labeled by the Carter constant, with other parameters and constants of motion fixed. Different invariant curves with different Carter constants are nested within each other on the Poincar\'e surface of section.
The complete collection of resonant orbits with the same rational frequency ratio and the same Carter constant also forms a closed curve on the Poincar\'e surface of section. This is shown in Fig.~\ref{fig:PS3D_zeta0} in which we exhibit some phase space trajectories of the $1/2$-resonant orbits for the integrable subset of the metric of Eq.~\eqref{metric} (see the caption for more detailed setting). The black closed curve consists of the complete collection of resonant orbits with different initial conditions but the same rational frequency ratio and Carter constant. This closed curve is sandwiched by invariant curves with an irrational frequency ratio infinitesimally close to the rational one.

\begin{figure}
    \centering
    \includegraphics[scale=0.04]{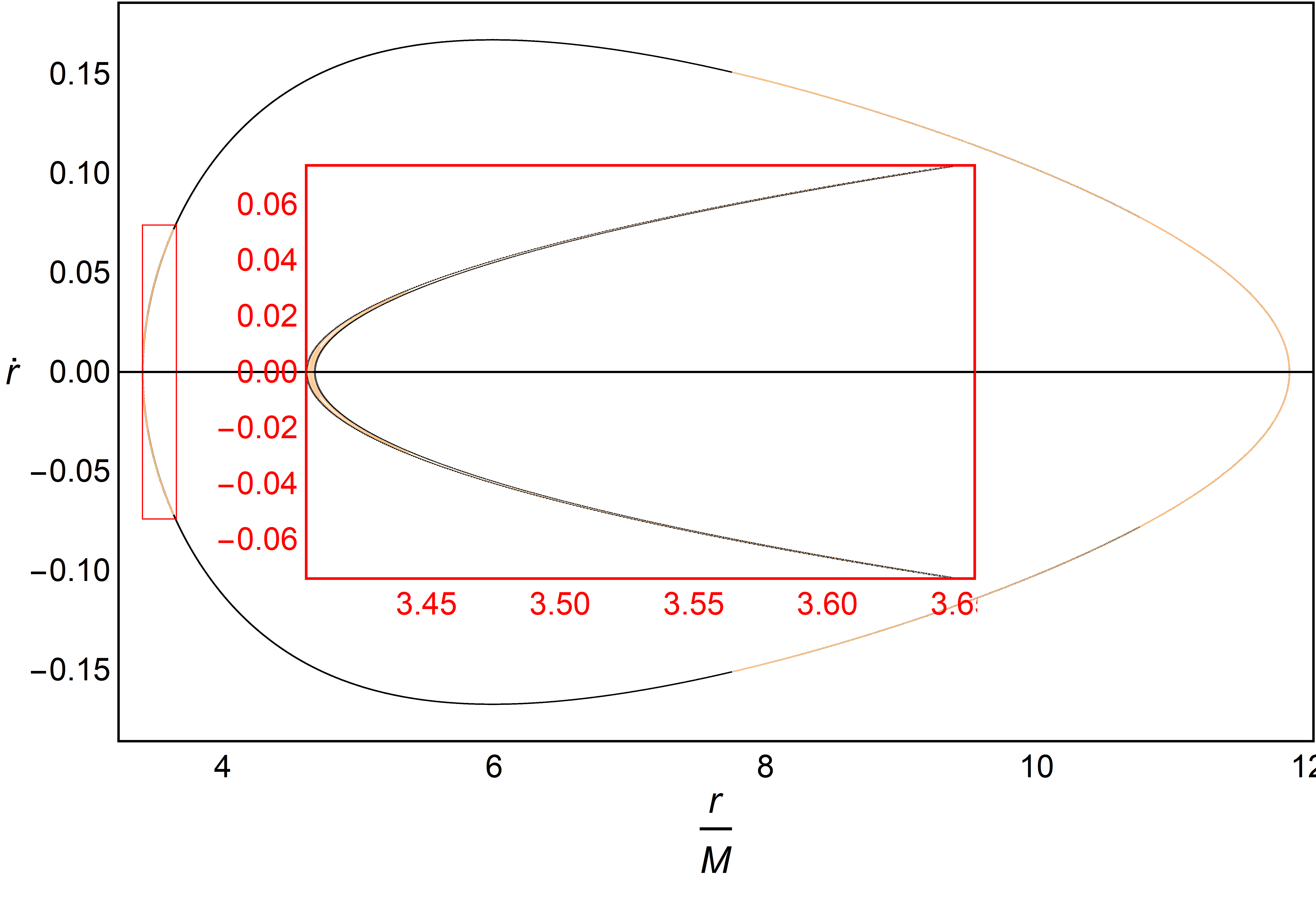}
    \caption{Poincar\'e surface of section at the equatorial plane ($\chi = 0$) of geodesics around the symmetric $1/2$-resonant Birkhoff island in the DSK spacetime \cite{Destounis:2020kss} with $\alpha_Q = -0.01$, $E=0.942$, $L_z=2.76M$, and $a=0.66M$. The yellow curves show the resonant orbits inside the Birkhoff island. The black curves that nearly sandwich the islands are the deformed KAM orbits.}
    \label{fig:sec_chi_0_KKn}
\end{figure}

Chaos may appear when an integrable system receives nonintegrable deformations, as in some cases of non-Kerr spacetimes, including the one described by Eq.~\eqref{metric}. Depending on the strength and features of the deformations, the chaos may leave nontrivial imprints on the Poincar\'e surface of section. In fact, according to the Kolmogorov-Arnold-Moser (KAM) theorem \cite{cbookKAM1,cbookKAM2}, if the deformations are small, the original invariant curves would be slightly deformed but remain continuous and nested. In this case, they are called KAM curves (black curves in Fig.~\ref{fig:sec_chi_0_KKn}).

On the other hand, even if the strength of chaos is small, according to the Poincar\'e-Birkhoff theorem \cite{cbookKAM1,cbookKAM3,cbookKAM4}, the resonant points on the Poincar\'e surface of section may be split into a series of periodic points. Half of the points are stable while the other half are not, and they distribute alternately on the Poincar\'e surface of section. The stable points attract and lock neighboring orbits whose frequency ratio is originally very close to the rational frequency ratio of the stable point. These orbits appear as a series of small islands surrounding the stable point called the Birkhoff chain of islands. Within each island, the orbits form a nested structure of closed curves. Because the island structure looms out of the resonant points, each chain of islands can be labeled by the order of resonance, which is precisely the frequency ratio of the radial and latitudinal oscillations of the resonant orbits, i.e., $\omega^r/\omega^\theta$. For example, we call the chain of islands in which the orbits have $\omega^r/\omega^\theta=1/2$ the $1/2$-resonant islands. Practically speaking, the existence of islands on the Poincar\'e surface of section is a clear signature of chaos. As an illustrative example, we show in Fig.~\ref{fig:sec_chi_0_KKn} the Poincar\'e surface of section and the structure of the $1/2$-resonant islands (yellow) in the case of a particular non-Kerr spacetime \cite{Destounis:2020kss} (from now on referred to as the DSK spacetime) different from the one in Eq.~\eqref{metric}. This non-Kerr spacetime has nonintegrable deformations controlled by a dimensionless parameter $\alpha_Q$ {\footnote{The non-Kerr metric proposed in Ref.~\cite{Destounis:2020kss} has the other deviation parameter $\alpha_{22}$. But this parameter does not break integrability. Therefore, we simply set $\alpha_{22}=0$ in this work.}}. We shall emphasize that although the DSK spacetime has chaotic orbital dynamics, it remains circular.

\subsection{Noncircularity, chaos, and the breakdown of the phase space discrete symmetry}
\label{sec:noncircularity_chaos}

The circularity condition and chaos are deeply intertwined. In Ref.~\cite{Wald:2010gr}, the author proved the circularity as the necessary and sufficient condition for the existence of the codimension-$2$ foliation by conserved quantities of the Killing vectors. Here we utilize the corollary of the proof and show that the nonexistence of the foliation not only gives rise to chaos but also leads to a particular kind of chaos different from those manifested in systems satisfying the circularity condition.

For the sake of completeness, let us briefly review the integrability in Frobenius' sense. When solving a geodesic system, the geodesic is integrable in the Liouville sense if enough conserved quantities exist. These conserved quantities become unique labeling for each geodesic, forming a foliation for the geodesic. The differences in these quantities between a pair of geodesics must remain the same under the geodesic evolution. As first investigated in Ref.~\cite{caviglia1982equation}, we may identify these quantities through the geodesic deviation equation
\begin{align}
\frac{d^2 s^\lambda}{d\tau^2} = R^\lambda_{\sigma\rho\eta} \frac{d x^\sigma}{d\tau} \frac{d x^\rho}{d\tau} s^\eta  \,,
\end{align}
that relates the evolution of the displacement between two geodesics $x^\mu(\tau)$ and $x^\mu(\tau)+s^\mu(\tau)$ to the curvature tensor $R^\lambda_{\sigma\mu\eta}$. For ``integrable'' orbits that are closed on the phase space, we should be able to express the geodesic deviation $s^\mu$ as a function of the position $x^\mu$ and the velocity $\dot x^\mu$ without explicit $\tau$ dependence. The inner product between the vector field $s$ and the velocity $\dot x$ would be a conserved quantity, as shown in Ref.~\cite{caviglia1982equation}.

Let us reorganize the equation using the Cartan connection as
\begin{align}
0 &= \frac{d^2 s^\lambda}{d\tau^2} - R^\lambda_{\sigma\rho\eta} \frac{d x^\sigma}{d\tau} \frac{d x^\rho}{d\tau} s^\eta  \nonumber\\
&= \dot x^\mu \dot x^\nu \nabla_\mu \nabla_\nu s^\lambda - R^\lambda_{\sigma\rho\eta} \dot x^\sigma \dot x^\rho s^\eta  \nonumber\\
&= \Omega_{\dot x}.s\,,
\end{align}
where $\Omega_{\dot x}$ is the curvature form of the connection $\omega_{\dot x} \equiv [\dot x, e]$, $[\cdot,\cdot]$ is the graded Lie bracket, $e$ is the vierbein, and $.$ denotes the inner product. This is precisely the Frobenius theorem that a set of commutative Lie vector fields along the geodesics implies the existence of conserved quantities foliating the phase space. The converse is trivial.

In a $d$-dimensional system with $d-2$ known commutative vector fields $V_1 \cdots V_{d-2}$, we may test if the last conserved quantity could exist by checking the commutativity of these $d-2$ vectors with the orthogonal directions. This leads to the circularity condition
\begin{align}
dV_j \land V_1 \land \dots \land V_{d-2} = 0 \,,\quad j = 1 \cdots d-2\,.
\end{align}

Apparently, the circularity condition is a unique realization of the Frobenius integrability in a system with $d-2$ conserved quantities. This setup is much broader than expected, as the most typical $d=4$ axisymmetric and stationary spacetime falls into this category. The corresponding conserved quantity is the Carter constant. Notice that while violating the circularity condition deprives us of the Liouville integrability, the reverse does not always hold. The last commutative vector field must be proportional to the orthogonal vector $* ( p \land V_1 \land \dots \land V_{d-2} )$ where $p$ is the conjugate momentum, and $*$ is the Hodge dual. If $\dot x$ explicitly depends on $\tau$, e.g., when considering a chaotic orbit, the circularity condition may be satisfied despite missing the final conserved quantity.

According to the Frobenius theorem, if the circularity condition is violated, the $d-2$ vector fields would no longer commute with their orthogonal vector field in the Lie derivative sense. In this case, the quantities associated with $d-2$ vector fields are still conserved, but any transformation along these vectors modifies the observable on the orthogonal subspace\footnote{The observable on the orthogonal subspace, e.g., $r$, $\chi$, $\dot r$ and $\dot\chi$ in the systems we are considering, must be generated by the reduced Hamiltonian on the orthogonal subspace and the orthogonal vector field.}. To wit, the system becomes non-abelian. It, therefore, bears both theoretical and practical interests to analyze what happens if we apply some transformations along those $d-2$ directions in a noncircular spacetime with additional discrete symmetries on the orthogonal subspace that shall not manifest themselves.

Let us first focus on the circular spacetimes, e.g., the DSK spacetime with a nonzero $\alpha_Q$ in the Boyer-Lindquist coordinates. Consider the discrete symmetries of the orthogonal subspace. In this system, there exists the equatorial-plane reflection symmetry ($\chi \to -\chi$, $\dot\chi \to -\dot\chi$) \footnote{It is possible to construct a stationary and axisymmetric metric that deviates from Kerr one in a way that breaks the equatorial-plane reflection symmetry. See Refs.~\cite{Cunha:2018uzc,Chen:2020aix,Lima:2021las} in which the orbital dynamics have no chaos and Refs.~\cite{Cardoso:2018ptl,Cano:2019ore} in which chaos may be present.}. In addition, we also have ($t,\varphi_\textrm{BL}$) simultaneous reversal symmetry \cite{Carter:1969zz}. By the Frobenius theorem, two symmetry transformations disentangle from each other. Therefore if we consider a simultaneous reversal of $(\tau,t,\chi,\varphi_\textrm{BL})$, we have ($\chi \to -\chi$, $\dot r \to -\dot r$), and the resulting changes in the conserved quantities $E$ and $L_z$ vanish identically. The Poincar\'e section of $(r,\dot{r})$ at the equatorial plane, once including both directions of piercings, must be $\dot r$-reflection-symmetric unless the geodesic of interest is unbounded (see Fig.~\ref{fig:sec_chi_0_KKn} and more in Ref.~\cite{Destounis:2020kss} for the symmetric islands in the case of the DSK spacetimes). Similarly, the Poincar\'e sections of $(\chi,\dot r)$ would be $(\chi,\dot r)$-simultaneous-reflection-symmetric.

On the other hand, in a system that breaks the circularity condition, the conserved quantities associated with the Killing vectors do not properly foliate the phase space. In the case of the noncircular spacetime of Eq.~\eqref{metric}, the $\chi$ reflection symmetry does not commute with $(u,\varphi)$ simultaneous reflection symmetry. Therefore, the Poincar\'e section of $(r,\dot r)$ at the equatorial plane would be asymmetric if the geodesic of interest does not explicitly possess $\chi$ reflection symmetry, i.e., when the geodesic is inside resonant islands. This observation suggests that the location of resonant islands is a probe of the circularity condition. The Poincar\'e sections of $(\chi,\dot r)$ would be $(\chi,\dot r)$-simultaneous-reflection-asymmetric as well.

With the breakdown of the symmetry, one may wonder why should the equatorial plane remains the natural choice for the Poincar\'e surface of section. We emphasize that the $\chi$ reflection symmetry still remains intact in the spacetime structure. Thus, if we focus on the latitudinal motion, the orbit must be symmetric to, and thus likely passing through, the equatorial plane
, making it the optimal Poincar\'e surface of section.

The subtle and partial breakdown of the symmetry is the salient feature of the noncircularity. In the next section, we will demonstrate explicitly through numerical treatments how the location of islands is sensitive to noncircular deformations, regardless of the strength of the deformations.



\section{Asymmetric Birkhoff islands}\label{sec.numerical}

In the previous section, we have laid down our arguments regarding possible chaotic signatures of orbital dynamics in noncircular spacetimes. In particular, some novel features may appear, such as asymmetric locations and structures of resonant islands. Here, we will demonstrate these features by directly showing the resonant orbits on the Poincar\'e surface of section. Considering the non-Kerr metric of Eq.~\eqref{metric}, we numerically solve the equations of motion of the bound orbital system by fixing some parameters while varying the rest to identify features emerging from a family of geodesics. We fix $E=0.942$, $L_z=2.76M$, and $a=0.66M$ for the rest of our paper. While solving the trajectories, we keep the accuracy of the Hamiltonian constraint down to $10^{-8}$ during the integration time. This is achieved by adopting the standard eighth-order implicit Runge-Kutta method. In the calculation of each trajectory, we terminate the calculation only after we can distinguish KAM curves from the closed curves inside islands.    

In Fig.~\ref{fig:islandsandh}, we choose $\zeta=1$ and $l_\textrm{NP}=0.4M$, then focus on the left branch of the $1/2$-resonant islands. The existence of Birkhoff islands on the Poincar\'e surface of section is confirmed, indicating that the orbital dynamics of the non-Kerr spacetime of Eq.~\eqref{metric} is chaotic. In particular, the distribution of Birkhoff islands on the Poincar\'e surface of section is not symmetric with respect to the horizontal axis ($\dot{r}=0$). For circular spacetimes with nonintegrable orbital dynamics, the resonant islands can only either be symmetric (see Fig.~\ref{fig:sec_chi_0_KKn}) or mirror images of down-piercing islands through $\dot\chi$-$\dot r$ inversion symmetry, as will be shown later in sec.~\ref{sec.pertur}. In the latter case, no island in the chain ever crosses the $\dot{r}=0$ axis. On the contrary, in noncircular spacetimes, as depicted in Fig.~\ref{fig:islandsandh}, the major portion of the chain of islands has $\dot{r}>0$ while a smaller but finite portion has $\dot{r}<0$. Such vertically asymmetric distribution of islands that also cross the $\dot{r}=0$ axis, is a feature of noncircularity.

\subsection{Quantify the asymmetry: Critical velocity}

\begin{figure}[!t]
    \centering
\includegraphics[scale=0.33]{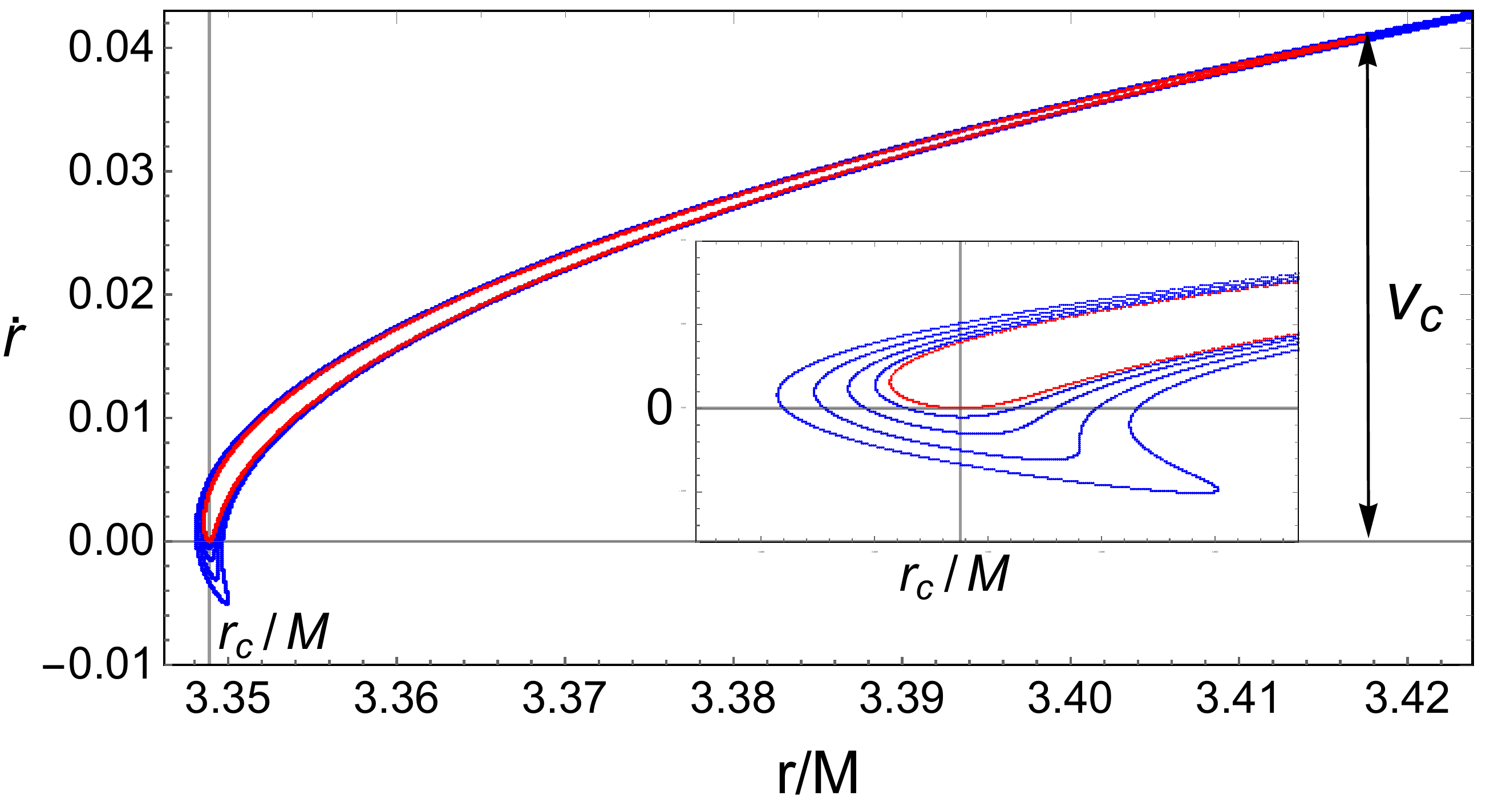}  \caption{Some orbits nested within the left branch of the $1/2$-resonant islands are shown. The islands are asymmetric with respect to the $\dot{r}=0$ axis. The critical velocity $\textrm{v}_c$ is defined on the orbit (red) that only touches the $\dot{r}=0$ axis at a single piercing $(r_c,0)$. The inset zooms in on the region near that piercing. In this figure, we choose $\zeta=1$ and $l_\textrm{NP}=0.4M$.} \label{fig:islandsandh}
\end{figure}

\begin{figure}[t]
    \centering
\includegraphics[scale=0.47]{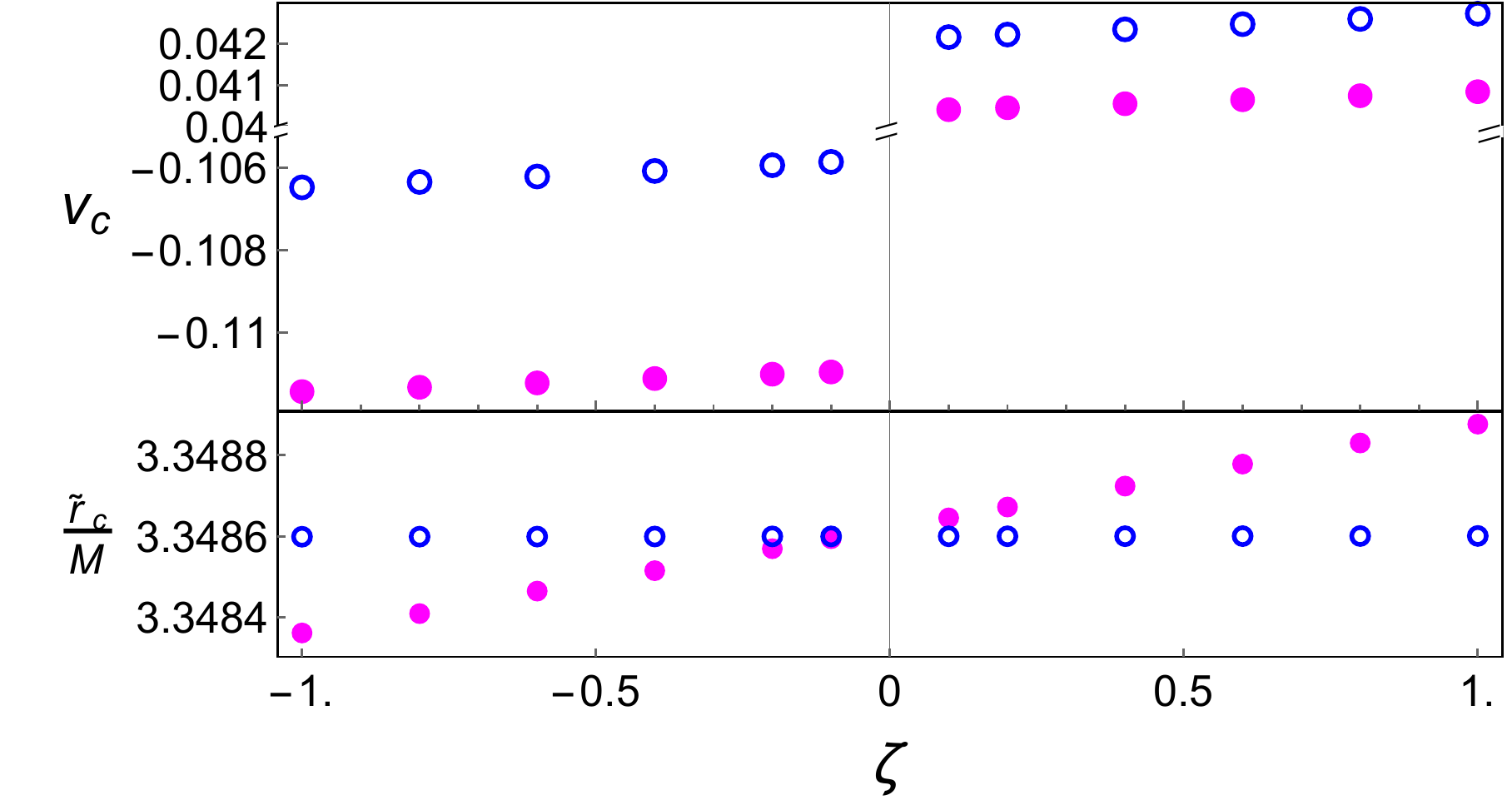}
    \caption{The values of $\tilde{r}_c$ and critical velocity $\textrm{v}_c$ with respect to $\zeta$ are shown. The blue circles and the magenta points correspond to $l_\textrm{NP}=0.1M$ and $0.4M$, respectively. The discontinuous jump of $\textrm{v}_c$ at $\zeta=0$ is clearly visible, although $\tilde{r}_c$ varies smoothly there.}   \label{fig:hpointzeta}
\end{figure}

To quantify how much asymmetrically the islands distribute on the Poincar\'e surface of section, we define the critical velocity $\textrm{v}_c$ as follows. We first consider the left branch of the $1/2$-resonant islands. Among the nested orbits within the islands, we look for the one that only touches the horizontal axis once, say, at $(r_c,0)$ (see the red contour in Fig.~\ref{fig:islandsandh}). Among the piercings recorded on the contour, we identify the one that has the maximum $|\dot{r}|$. Then we define the critical velocity $\textrm{v}_c$ as the $\dot{r}$ of that piercing (see Fig.~\ref{fig:islandsandh} for graphical illustrations). If the islands are symmetric with respect to the horizontal axis, the critical velocity vanishes, and the associated orbit is precisely the central point within the islands. In Fig.~\ref{fig:hpointzeta}, we consider two values of $l_\textrm{NP}=0.1M$ (blue) and $0.4M$ (magenta), with other parameters $(a,E,L_z)$ fixed, and show how the critical velocity $\textrm{v}_c$ and a shifted version of $r_c$, i.e., $\tilde{r}_c\equiv r_c+18(5l_\textrm{NP}-2M)/625$\footnote{ $\tilde r_c$ is constructed such that the shift vanishes when $l_\textrm{NP} = 0.4M$, relieving us from this artificial shift for the rest of the paper.}, vary with respect to $\zeta$. Interestingly, even though $\tilde{r}_c$ changes smoothly, the critical velocity has a discontinuous jump at $\zeta=0$ where the orbital dynamics is integrable. Furthermore, when approaching the Kerr limit (blue), the critical velocity and $\tilde{r}_c$ become less sensitive to $\zeta$, but $\textrm{v}_c$ remains sizable, and its discontinuity at $\zeta=0$ remains.

\subsection{Quantify the asymmetry: Latitude of symmetry}

\begin{figure}[!ht]
    \centering
\includegraphics[scale=0.38]{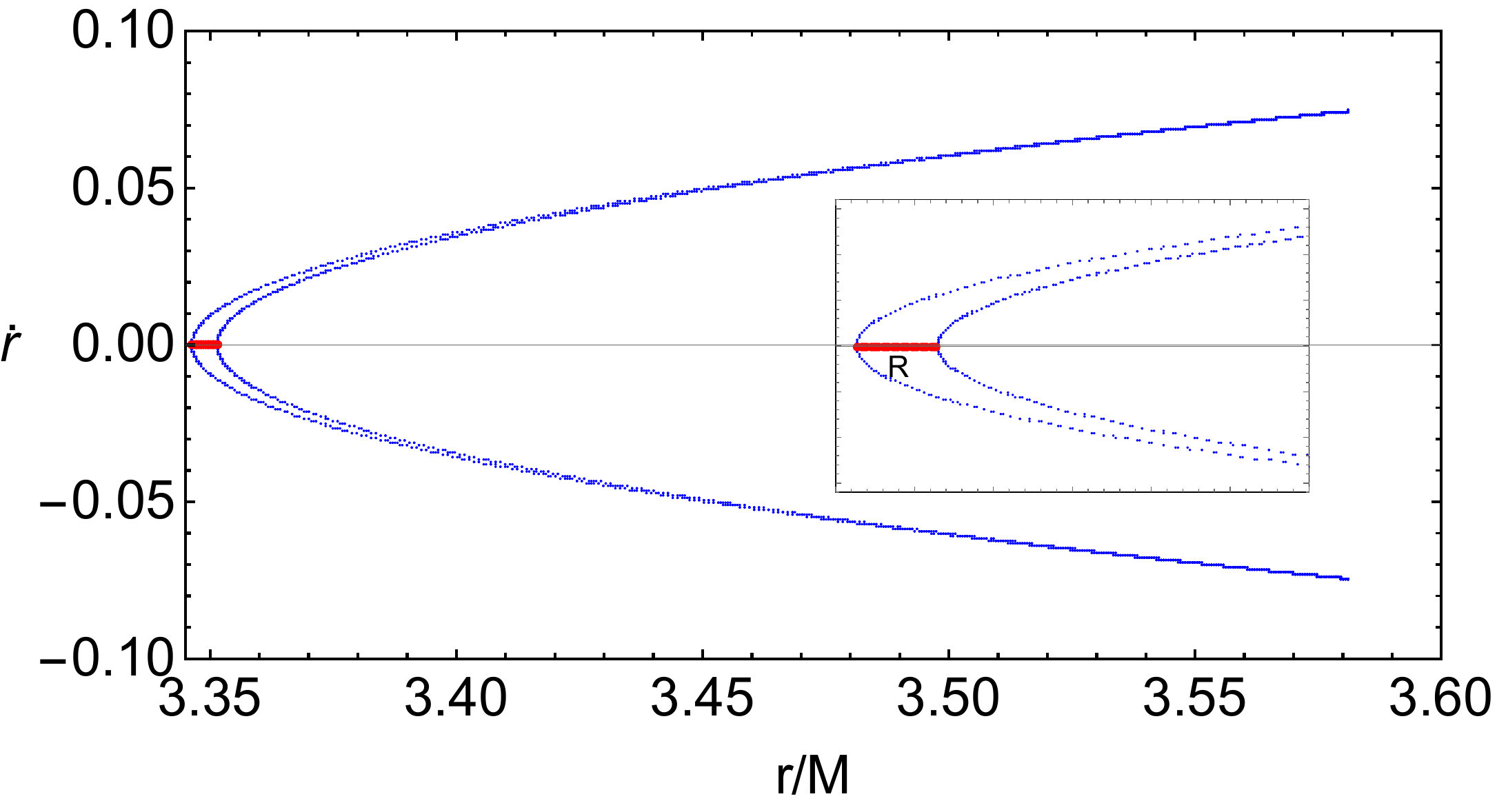}
    \caption{We show the Poincar\'e surface sectioned at $\chi=\chi_s=-0.15064$. In this figure, the values of ($E,L_z,a,\zeta,l_\textrm{NP}$) are the same as those in Fig.~\ref{fig:islandsandh}. The blue contour is the outermost orbit among those nested ones inside the $1/2$-resonant island (The initial condition for this orbit is $r(0)=3.3514586M$, $\chi(0)=\chi_s$ and $\dot{r}(0)=0$). The inset zooms in on the island around $\dot{r}=0$. The line segment in red represents the horizontal width $R$ of the island.}
    \label{fig:SymmetricIsland}
\end{figure}

Though the standard Poincar\'e surface of section is usually chosen as the surface that intersects orbits at $\chi=0$ for the reason mentioned in the last section, one can in principle consider the Poincar\'e surface at any $\chi=\textrm{constant}$ as long as every single orbit inside the island on the equatorial plane still intersect the new Poincar\'e surface for every single revolution. In Fig.~\ref{fig:islandsandh}, we see that, inside the Birkhoff islands, the piercings on the $\chi=0$ plane are asymmetric with respect to the horizontal axis ($\dot{r}=0$). This indicates that for a resonant orbit on the noncircular spacetime of Eq.~\eqref{metric}, the maximum radial velocity at $\chi=0$ cannot be simply expressed as  $\pm|\dot{r}_{\textrm{max}}|$ during its whole evolution. This asymmetry varies as we shift the section along the latitudinal direction. In particular, it can be completely removed at a particular value of $\chi$, given a set of metric parameters. We dub it the latitude of symmetry, denoted as $\chi_s$. In Fig.~\ref{fig:SymmetricIsland}, the island appears symmetric on the Poincar\'e surface of section that intersects the phase space at $\chi=\chi_s$. For simplicity of the presentation, we show only one orbit in the $1/2$-resonant islands in Fig.~\ref{fig:SymmetricIsland}, but every orbit with the same set of $(a,\zeta,l_\textrm{NP})$ is symmetric with respect to the $\dot{r}=0$ axis. In this figure, the parameters $(E,L_z,a,\zeta,l_\textrm{NP})$ are fixed as those in Fig.~\ref{fig:islandsandh}, and the latitude of symmetry $\chi_s$ is $\chi_s=-0.15064$.

At this point, we have shown that, if the spacetime is circular, the Birkhoff islands and the stable points, around which the resonant orbits are nesting, could appear symmetric with respect to the $\dot{r}=0$ axis on the Poincar\'e surface sectioned at $\chi=0$, as one can see from Fig.~\ref{fig:sec_chi_0_KKn}. In this case, the piercings of orbits would have $\dot{r}$ ranging from $+|\dot{r}_{\textrm{max}}|$ to $-|\dot{r}_{\textrm{max}}|$ on that surface of section.\footnote{Later, we will show that the Birkhoff islands in some cases of circular spacetimes, e.g. the DSK metric with $\alpha_Q>0$, could not even appear symmetric on $(r,\dot{r})$ surface sectioned at any value of $\chi$. There does not exist any $\chi_s$ at which the islands appear symmetric. Rather they would be symmetric on the section of $\dot\chi = 0$, or on the latitude where $\dot\chi$ vanishes for the central periodic orbit of the island if we include both up and down piercing of the Poincar\'e surfaces of section.} On the other hand, if the spacetime is noncircular, like the one considered in this section, the Birkhoff islands and stable points on the Poincar\'e surfaces of section at $\chi=0$ and $\dot\chi = 0$ are generically not symmetric with respect to the $\dot{r}=0$ axis. The resonant orbits inside the islands and KAM curves can appear symmetric with respect to the $\dot{r}=0$ axis only when the surface of $\chi=\chi_s$ is chosen. It implies that the radial turning points ($\dot r = 0$) of the orbit corresponding to stable fixed points, i.e., the center of nested curves inside islands, move away from the equatorial plane and instead reside on the $\chi=\chi_s$ plane.

\begin{figure}[!ht]
    \centering
\includegraphics[scale=0.39]{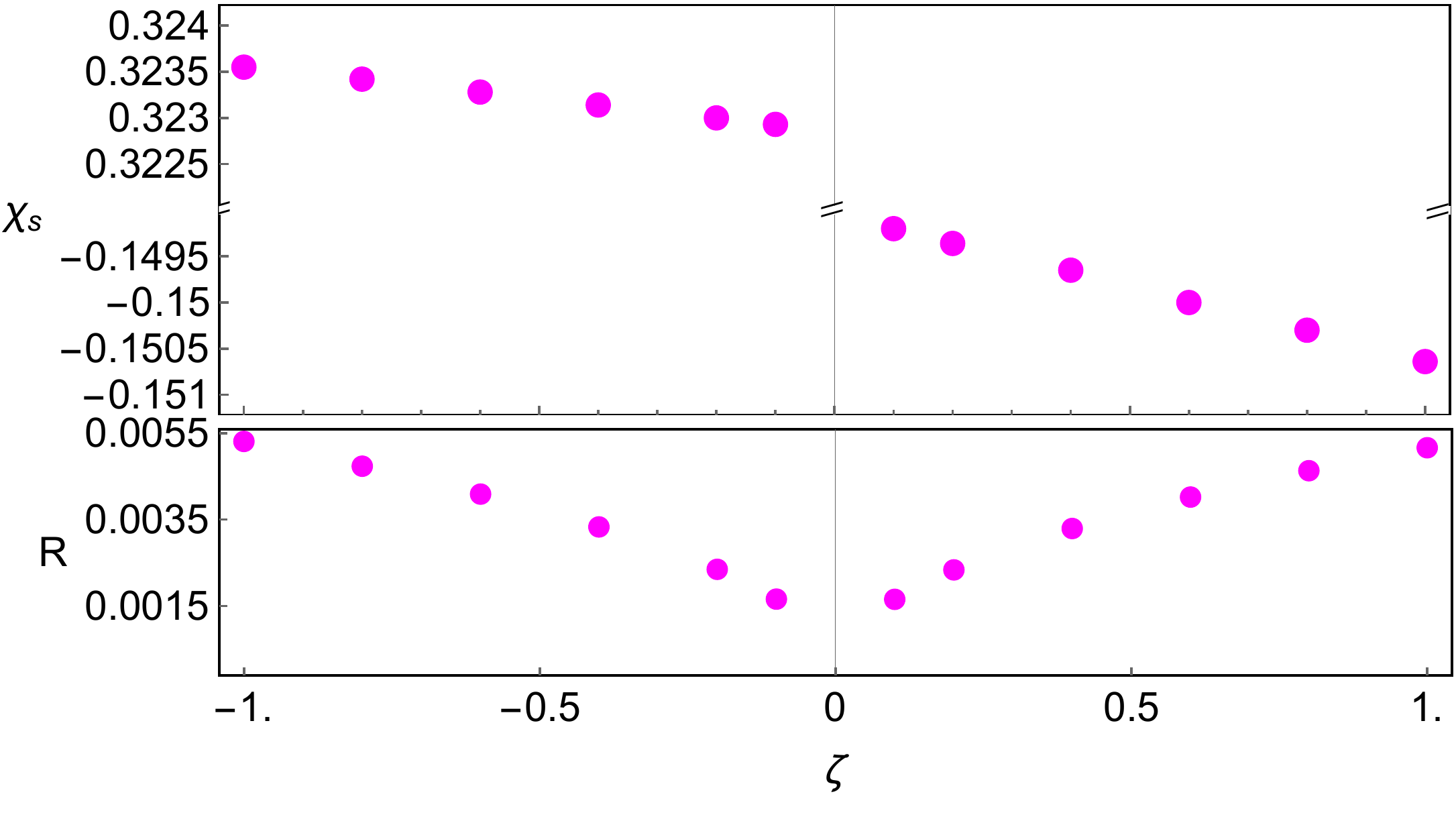}
    \caption{The latitude of symmetry $\chi_s$ and the width of island $R$ are shown with respect to $\zeta$, with other parameters fixed as those in Fig.~\ref{fig:islandsandh}. A discontinuous jump of $\chi_s$ across $\zeta=0$ is clearly visible, as that of $\textrm{v}_c$ in Fig.~\ref{fig:hpointzeta}. On the other hand, the width of islands as defined by the red line in Fig.~\ref{fig:SymmetricIsland} decreases with $|\zeta|$ and approaches zero as $|\zeta|\rightarrow 0$. The width scales almost perfectly as $R \propto |\zeta|^{1/2}$, as predicted (c.f. sec.~2.6.4 of Ref.~\cite{cbookKAM1}, sec.~6.2.2 of Ref.~\cite{cbookKAM4} and Ref.~\cite{Lukes-Gerakopoulos:2021ybx}). }
    \label{fig:R_Vs_zeta}
\end{figure}

In fact, the introduction of the latitude of symmetry $\chi_s$ has its own unique merit. As we have mentioned, given a set of $(a,\zeta,l_\textrm{NP})$, each orbit on the Poincar\'e surface of section at $\chi=\chi_s$ is symmetric with respect to the $\dot{r}=0$ axis. Because $\chi_s$ is uniquely determined in this sense, it can serve as a measure of the asymmetry of islands that could be induced by noncircularity, just as the critical velocity $\textrm{v}_c$. For example, for the DSK spacetime with a negative $\alpha_Q$, these two measures, $\textrm{v}_c$ and $\chi_s$, always vanish. In Fig.~\ref{fig:R_Vs_zeta}, we fix $l_\textrm{NP} = 0.4M$ as in Fig.~\ref{fig:islandsandh} and show how $\chi_s$ varies with $\zeta$. Similar to what we have observed in $\textrm{v}_c$, a discontinuous jump at $\zeta=0$ is clearly visible.

The symmetric pattern on the Poincar\'e surface of section at $\chi=\chi_s$ also allows us to define the width of the islands. As one can see in Fig.~\ref{fig:SymmetricIsland}, we first consider the outermost orbit within the left branch of the $1/2$-resonant islands. Since the pattern appears symmetric with respect to the $\dot{r}=0$ axis, one can unambiguously define the width of the island as the line segment connecting the two piercings of the orbit on the $\dot{r}=0$ axis (see the red line segment in Fig.~\ref{fig:SymmetricIsland}). Essentially, the width of island, $R$, quantifies the strength of chaos when varying the nonintegrable deformation parameters of the system. In the bottom panel of Fig.~\ref{fig:R_Vs_zeta}, we show how the width $R$ varies with respect to $\zeta$. One can see that it approaches zero from both sides of the $\zeta=0$ axis. This is expected because the orbital dynamics is integrable at $\zeta=0$. Furthermore, the non-KAM-curve region on the frequency ratio map, which can be estimated by the width of islands, scales at the rate of the square root of the nonintegrable deformation parameter \cite{cbookKAM1}. The width $R$ follows closely with this rule all the way up to $\zeta = \pm 1$. This suggests the possibility of using perturbation analysis in the parameter space to understand the origin of the asymmetric features.

At this stage, we can draw the following conclusions according to our analysis. Although the strength of chaos smoothly reduces to zero at the integrable limit, i.e., $|\zeta|\rightarrow0$, the asymmetry in the distribution of Birkhoff islands on the Poincar\'e surfaces of section does not. Also, the central $1/2$-resonant orbit tends to move away from the equatorial plane such that the whole family of $1/2$-resonant islands emerges asymmetrically on the Poincar\'e section. We will see in the next section that this can be interpreted as the existence of an off-equatorial net force on the orbit. The asymmetric distribution of islands is an important feature of the orbital dynamics in noncircular spacetimes.

In sec.~\ref{sec:noncircularity_chaos}, we have briefly mentioned that because the conserved quantities associated with the Killing vectors do not properly foliate the phase space, the Birkhoff islands could have asymmetric distribution on the Poincar\'e surfaces of section for noncircular spacetimes. This is then supported by the numerical analysis in sec.~\ref{sec.numerical}. In the next section, by using perturbation analysis, we will dig slightly deeper to understand, from an analytic point of view, why the islands acquire such patterns and why the asymmetry appears to be changing discontinuously with respect to parameter shifts at the integrable limit.


\section{Perturbation analysis of the island hamiltonian}\label{sec.pertur}

As we have seen in the previous section, the resonant islands are no longer $\dot r$-symmetric on the equatorial plane for the noncircular spacetime of Eq.~\eqref{metric}. Furthermore, such asymmetry appears to depend strongly on the form of noncircularity rather than on its strength. The location of islands thus becomes a highly sensitive probe for noncircularity. To better characterize the relation between the two, we utilize the perturbation analysis to extract the quantitative description of the orbits.

Let us expand the chaotic geodesic system on a stationary, axisymmetric manifold $g$ around an integrable system on another stationary, axisymmetric manifold $g_{(0)}$. Since the orbits on $g_{(0)}$ are integrable, there is a hidden conserved quantity, say, $C$. We will denote the background objects by subscript $(0)$ and the perturbative parts by subscript $(1)$. The Hamiltonian and the equations for the conserved quantities of the background system
\begin{align}
-\frac{1}{2} &= H_{(0)} = \frac{1}{2} p_{(0), \mu} g_{(0)}^{\mu\nu} p_{(0), \nu}  \label{eq:constraint_H}  \,,\\
-E &\equiv p_{(0),u} =  g_{(0), u\nu} \dot x_{(0)}^\nu  \,,\\
L_z  &\equiv p_{(0),\varphi} =  g_{(0), \varphi\nu} \dot x_{(0)}^\nu  \,,\\
\dot C_{(0)} &= \left\{ H_{(0)} , C  \right\}_\text{P.B.} =0  \,,  \label{eq:constraint_C}
\end{align}
determine exactly the geodesics on the background manifold $x_{0}^\mu (\tau)$ as functions of the proper time $\tau$. $\left\{\right\}_\text{P.B.}$ in Eq.~\eqref{eq:constraint_C} is the Poisson bracket. The equations of the perturbed system can be expressed as
\begin{align}
0 \approx H_{(1)} &\approx  p_{(1), \mu} g_{(0)}^{\mu\nu} p_{(0), \nu} + \frac{1}{2} p_{(0), \mu} g_{(1)}^{\mu\nu} p_{(0), \nu}  \,,\\
0 = p_{(1),u} &\approx  g_{(1), u\nu} \dot x_{(0)}^\nu +  g_{(0), 0\nu} \dot x_{(1)}^\nu  \,,\\
0 = p_{(1),\varphi} &\approx  g_{(1), \varphi\nu} \dot x_{(0)}^\nu +  g_{(0), 0\nu} \dot x_{(1)}^\nu  \,,
\end{align}
where only $\dot r_{(1)}$ and $\dot \chi_{(1)}$ remain to be solved. Notice that the perturbation to the constants of motion $\mu$, $E$, and $L_z$ vanishes. In other words, as we introduce the perturbation, the initial condition $x^\mu$ has to vary consistently to ensure that the constants of motion remain unperturbed. This is natural for our methodology in sec.~\ref{sec.numerical} as we fix the constants of motion regardless of the metric deformations \footnote{One may take different approaches to the issue of the initial condition. For instance, if the orbit is fitted against observational data, the observed orbital frequency and the impact parameter may be a better pair to be fixed than $E$ and $L_z$.}.

In general, such a system is chaotic, as we have seen in previous sections. However, at the center of the Birkhoff island lies a stable fixed point, that corresponds to a recurring geodesic with a short period. Let us attempt to locate that geodesic perturbatively.

Since the orbit is closed, the first-order perturbation to the hidden conserved quantity $C$ must be periodic. We may evaluate its change rate along the perturbed orbit by plugging it into the first-order Hamilton-Jacobi equation. The change rate $\dot C_{(1)} = \left\{ H_{(1)}, C \right\}_\text{P.B.}$ contains only explicit first-order terms as $C$ is conserved at the zeroth order. We may therefore subject it to the background geodesic with the Hamiltonian constraint Eq.~\eqref{eq:constraint_H}. 

Let us be more specific about the quantity $C$. Here we are considering a particular form of the Carter constant
\begin{align}
C &= a^2  \chi^2 + K^{\mu\nu} p_\mu p_\nu  \label{eq:Carter}  \,,\\
K^{uu} &= a^2 \left(1-\chi ^2\right)  \,,\,\,
K^{u\varphi} = 2a  \,,\,\,
K^{\varphi\varphi} = \left(1-\chi ^2\right)^{-1}  \,,  \nonumber\\
K^{\chi\chi} &= \left(1-\chi ^2\right)  \,,\,\,
K^{r\mu} = K^{u\chi} = K^{\chi\varphi} = 0  \,,
\end{align}
where $K^{\mu\nu}$ is related to the Killing tensor of the background spacetime $\mathcal{K}^{\mu\nu}$ via $\mathcal{K}^{\mu\nu} \equiv K^{\mu\nu} -a^2 \chi^2 g^{\mu\nu}$. The Carter constant $C$ foliates the phase space of a large class of spacetimes, including the spacetime described by Eq.~\eqref{metric} with $\zeta$ vanishing identically and, of course, the Kerr spacetime.


We may now write down the first-order change rate of the Carter ``constant'' in terms of $r$, $\chi$, and $C$ as
\begin{widetext}
\begin{align}
\dot C_{(1)} &= 576 \zeta a^2 M M_{\zeta \to 0}^2 l_\textrm{NP}^4 r^{-5} \chi  \text{ sign} \left( \dot\chi \right) \sqrt{ \left( 1-\chi^2 \right) \left( C - a^2 \chi^2 - a^2 \left( 1-\chi^2 \right) E^2 + 2 a E L_z \right) - L_z^2 }  \nonumber\\
&\times \left( \left( \left( r^2 + a^2 \right) E - a L_z \right) + \text{sign}\left( \dot r \right) \sqrt{ \left( \left( r^2 + a^2 \right) E - a L_z \right)^2 - \left( r (r-2M_{\zeta \to 0})+ a^2 \right)  \left( C + r^2 \right) } \right)^2  \nonumber\\
&\times \left( r (r - 2 M_{\zeta \to 0}) + a^2 \right)^{-2} \left( r^2 + a^2 \chi ^2 \right)^{-2}  \label{eq:dotC_AH}  \,,
\end{align}
for the metric of Eq.~\eqref{metric}, with $M_{\zeta \to 0}$ denoting the mass function in Eq.~\eqref{eq:mass_func} at $\zeta \to 0$ limit. Similarly, we can also obtain the change rate for the DSK metric:
\begin{align}
\dot C_{(1)} &= 2M^3 \alpha_Q a^2 \chi  \text{ sign} \left( \dot\chi \right) \sqrt{ \left( 1-\chi^2 \right) \left( C - a^2 \chi^2 - a^2 \left( 1-\chi^2 \right) E^2 + 2 a E L_z \right) -L_z^2 }  \nonumber\\
&\times \bigg{(} \left( \left( \left( r (r-2M) + a^2 \right) \left( r^2 + a^2 \chi^2 \right) + 4M r \left( r^2 + a^2 \left( 1-\chi^2 \right) \right) \right) E - 4M r a L_z \right)^2 -4M^2 r^2 \left( \left( r^2 + a^2 \right) E - a L_z \right)^2  \nonumber\\
&- 2 \left( r (r-2M) + a^2 \right) \left( r^2+ a^2 \chi^2 \right) \left( \left( \left( r^2+a^2 \right) E - a L_z \right)^2 - \left( r (r-2M) + a^2 \right) \left( C+ r^2 \right) \right) \bigg{)}  \nonumber\\
&\times r^{-1} \left( r (r-2M) + a^2 \right)^{-2} \left( r^2 + a^2 \chi^2 \right)^{-4}  \,.\label{evolutionc1}
\end{align}
\end{widetext}

To ``separate'' the geodesic equations, we have utilized the background constraints
\begin{align}
\dot{r}^2 &= \left( r^2 + a^2 \chi^2 \right)^{-2} \Big{(} \left( \left( r^2 + a^2 \right) E - a L_z \right)^2  \nonumber\\
&- \left( r ( r - 2M_{\zeta \to 0} ) + a^2  \right) \left( C + r^2 \right) \Big{)}  \label{eq:r_cycle}  \,,\\
\dot{\chi}^2 &= \left( r^2 + a^2 \chi^2 \right)^{-2} \Big{(} \left( 1-\chi^2 \right) \big{(} C - a^2 \chi^2  \nonumber\\
&- a^2 \left( 1 - \chi^2 \right) E^2 + 2 a E L_z \big{)} - L_z^2 \Big{)}  \label{eq:chi_cycle}  \,,
\end{align}
for the metric described by Eq.~\eqref{metric} with $\zeta = 0$. For the DSK metric, the mass function $M_{\zeta \to 0}$ in the equations is replaced by the bare mass $M$. One may explicitly separate the radial and latitudinal motions by casting the radial equation of Eq.~\eqref{eq:r_cycle} into $dr/d\sigma=\pm\sqrt{V_r(r)}$, where $V_r$ is the radial potential and $\sigma$ is the Mino time. It is straightforward to identify the radial turning points ($\dot r = 0$) as solutions of vanishing $V_r$. One can apply the same procedure to the latitudinal equation and locate the latitudinal turning points ($\dot\chi = 0$). Furthermore, without specifying the initial condition, the conserved quantities $(E, L_z, C)$ alone can determine the period ratio between the radial and the latitudinal motion in Mino time, which must be the inverse of the frequency ratio in proper time. Therefore, we can identify the conserved quantities corresponding to a specific frequency ratio. For the $1/2$-resonant fixed point of the non-Kerr metric described by Eq.~\eqref{metric} with $\zeta = 0$ and $l_\textrm{NP} = 0.4M$, the latitudinal turning points are at $\chi = \pm 0.3555...$, and the radial turning points are at $r/M = 3.348...$ and $r/M = 11.933...$, respectively.

Notice that Eqs.~\eqref{eq:r_cycle} and \eqref{eq:chi_cycle} determine the background geodesic up to initial conditions $r(0)$, $\chi(0)$, and signs of $\dot r$ and $\dot\chi$. We can and will fix the orbit to start at the periapsis (minima of $r$) by shifting the proper time. For quasiperiodic orbits of the background system, given enough time, a single orbit would densely cover the $(r,\chi)$ torus. For resonant orbits, however, the latitude at the periapsis remains constant throughout the evolution, suggesting that each $\chi(0)$ leads to a different, non-overlapping orbit that appears as a different set of fixed points on the section. What we strive to investigate later is that, in the integrable limit of a perturbed system, which resonant orbit would correspond to the center of the $1/2$-resonant islands.

\begin{figure}
    \centering
    \subfigure[Metric of Eq.~\eqref{metric} with $l_{\textrm{NP}} = 0.4M$ and $\zeta = 1$]{\includegraphics[scale=0.081]{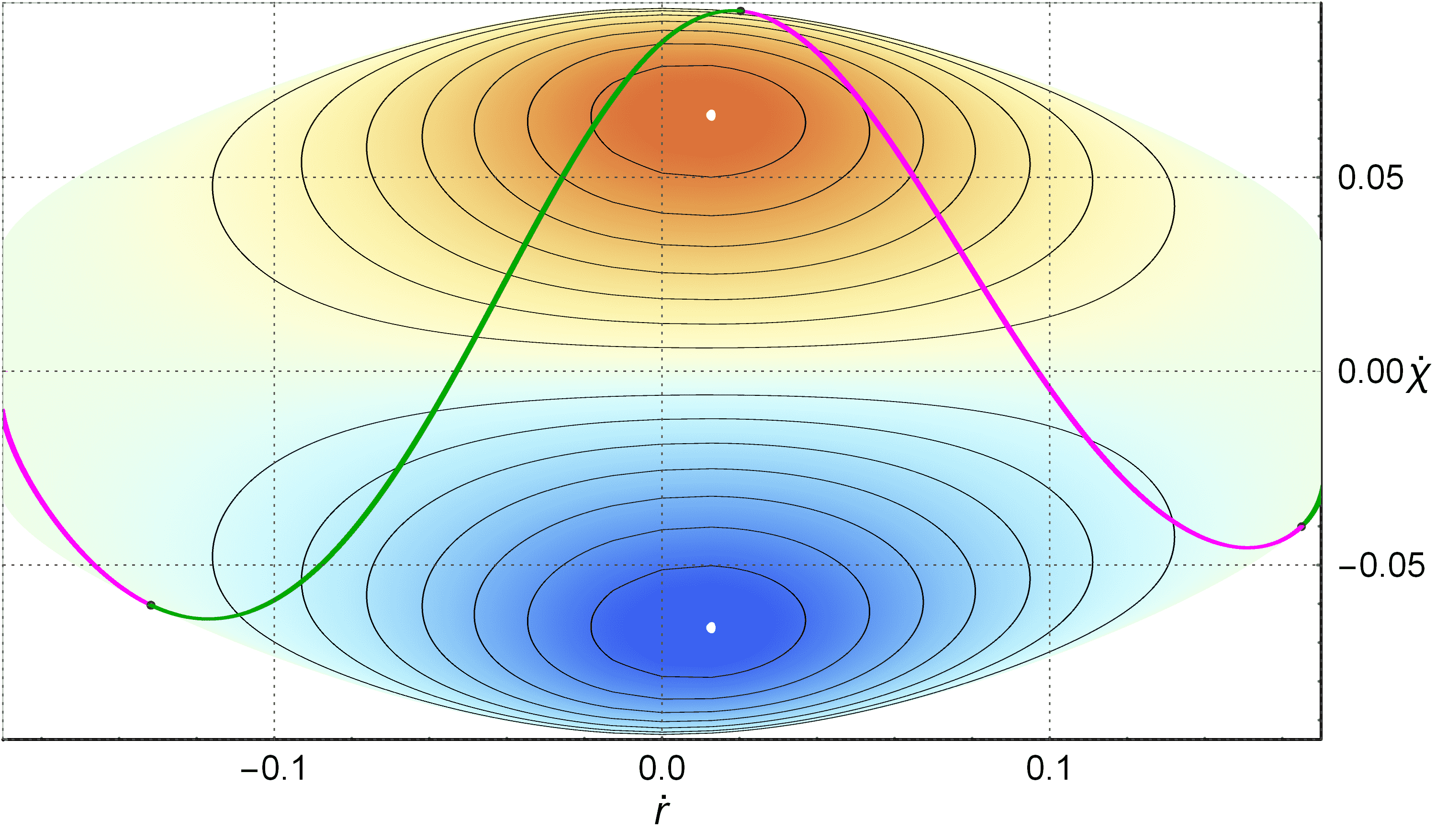}}
    \\
    \subfigure[DSK metric with $\alpha_Q = -0.01$]{\includegraphics[scale=0.081]{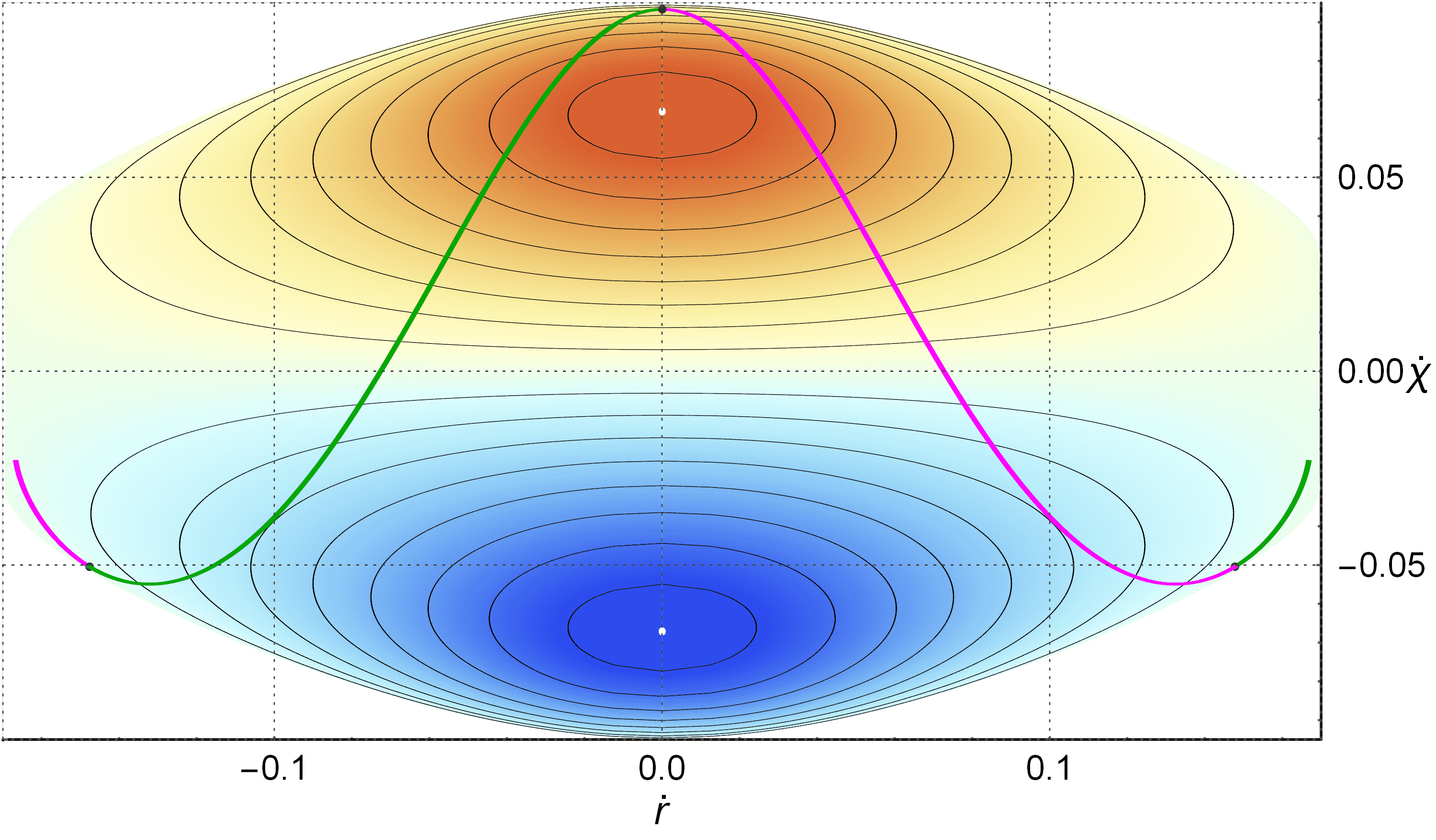}}
    \caption{The heat map (orange: positive, blue: negative, white: peak) of the change rate of the Carter ``constant'' $\dot C$ given by Eqs.~\eqref{eq:dotC_AH} and \eqref{evolutionc1}, with constant-spacing contours on $(\dot r,\dot\chi)$ plane. The results are evaluated using the orbit parameters of the geodesic at the center of the $1/2$-resonant islands considered in previous sections. The absolute value of $\dot C$ is irrelevant. The map is overlaid by the projection of the geodesic (green: $\chi < 0$, magenta: $\chi >0$) at the center of the island with the metric parameters given in the subcaption. $\dot C$ and the geodesic outside $r=6$ are cut as $\dot C$ is suppressed polynomially in terms of $r$. It is important to note that the sign of $\dot C$ flips as the geodesic reaches the equatorial plane (transition between green/magenta), and we only show the positive $\chi$ branch (magenta) of $\dot C$.
    \label{fig:DeltaC}
    }
\end{figure}

\begin{figure}
    \centering
    \subfigure[Metric of Eq.~\eqref{metric} with $l_{\textrm{NP}} = 0.4M$ and $\zeta = 1$]{\includegraphics[scale=0.081]{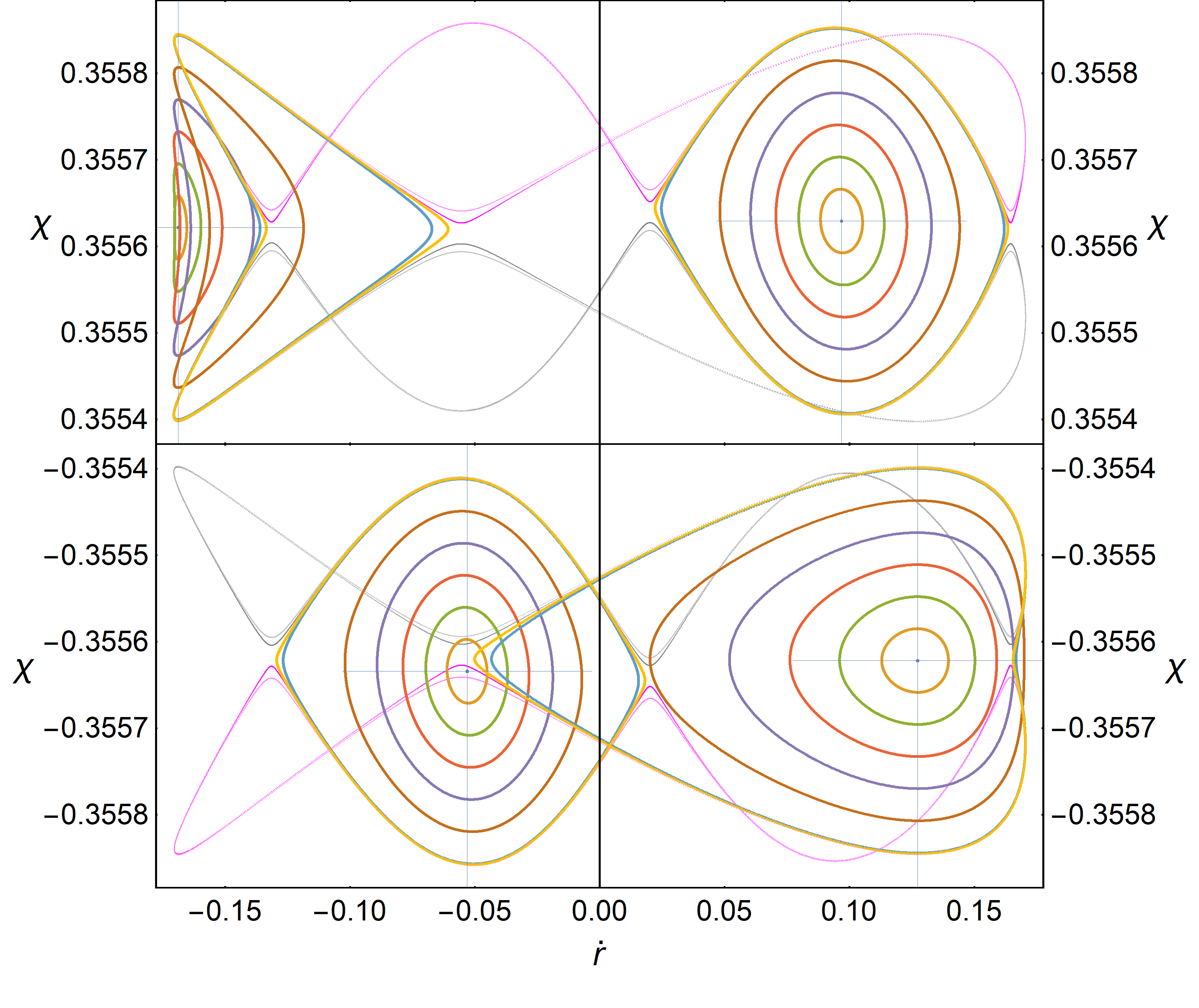}}
    \\
    \subfigure[DSK metric with $\alpha_Q = -0.01$]{\includegraphics[scale=0.081]{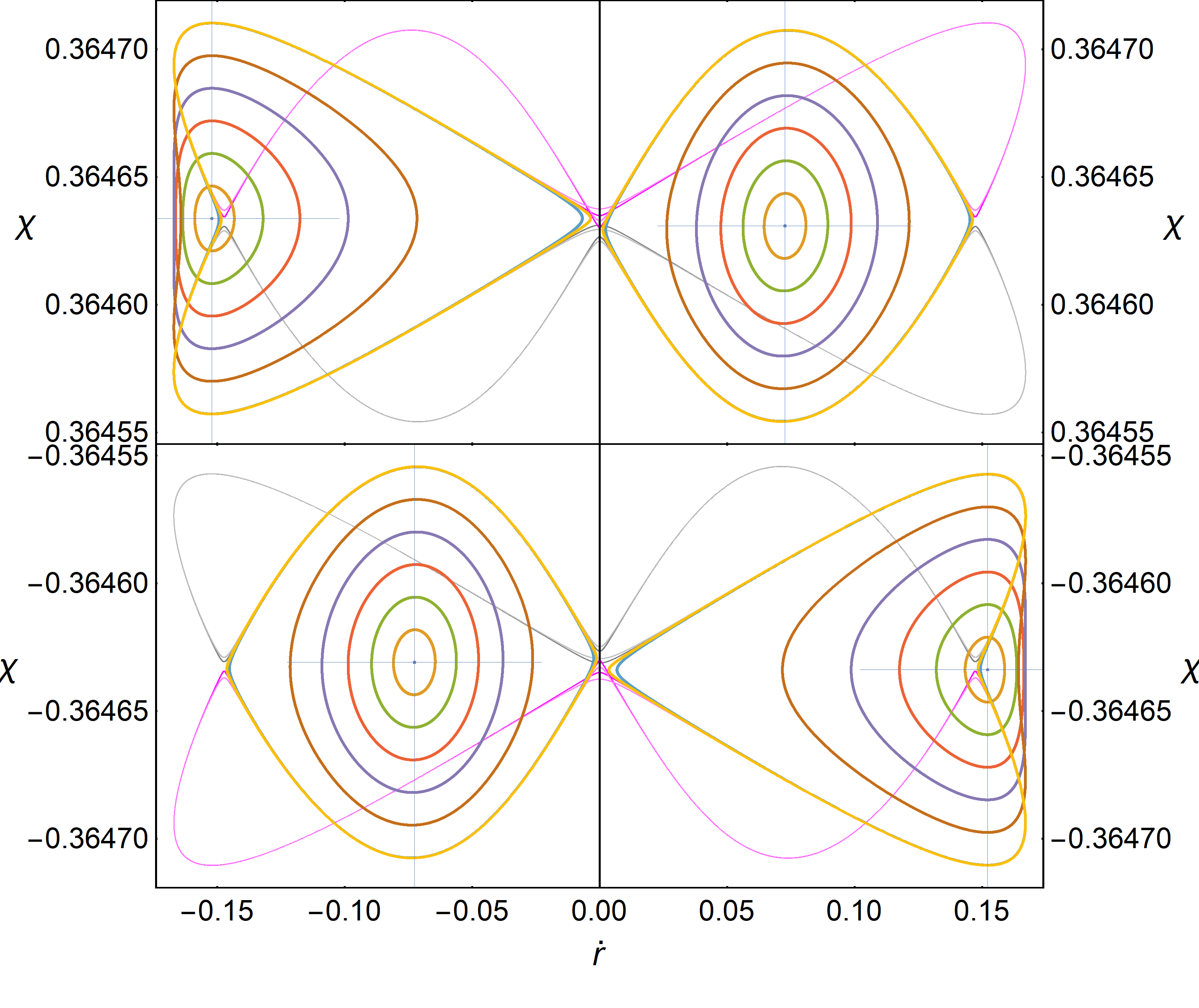}}
    \caption{The section of $\dot\chi = 0$, i.e., the latitudinal turning points, projected onto the $(\dot r,\chi)$ plane. Gray: The KAM curves with the radial frequency to latitudinal frequency ratio $\omega^r/\omega^\theta$ slightly smaller than $1/2$. Magenta: The KAM curves with $\omega^r/\omega^\theta$ slightly larger than $1/2$. Cross: The center of the island, i.e., the stable fixed point. Other circles: Geodesics from the edge of the island to the center of the island.}
    \label{fig:secs_chidot}
\end{figure}

Let us turn our attention back to the evolution of the Carter ``constant.'' At first glimpse, there are terms containing $\text{sign}\left( \dot r \right)$ in Eq.~\eqref{eq:dotC_AH}. Because of the presence of such terms, the evolution of $C$ is $\dot r$-inversion-asymmetric in the noncircular spacetime of Eq.~\eqref{metric}. This is the direct manifestation of the $(\chi,\dot r)$ asymmetry discussed in section~\ref{sec:noncircularity_chaos}. To better illustrate the asymmetry, we utilize Eqs.~\eqref{eq:r_cycle} and \eqref{eq:chi_cycle} to plot the change rate of the Carter ``constant'' against $\dot r$ and $\dot\chi$ in Fig.~\ref{fig:DeltaC}, alongside the numerically identified geodesic at the center of the resonant island. If we take the Poincar\'e section of $\dot\chi = 0$, the asymmetry of the Carter constant evolution deforms the radial evolution of Eq.~\eqref{eq:r_cycle} unevenly between $\dot r>0$ and $\dot r<0$ branches, breaking the $(\dot r,\chi)$ symmetry of the orbits, including KAM curves. The asymmetric patterns are visible in the top panel of Fig.~\ref{fig:secs_chidot}. For the DSK metric, on the other hand, the $(\dot{r},\chi)$ symmetry is preserved, as can be seen from the bottom panel.

\begin{figure*}
    \centering
    \subfigure[Metric of Eq.~\eqref{metric} with $l_\textrm{NP} = 0.4M$ and $\zeta = -1$]{\includegraphics[scale=0.084]{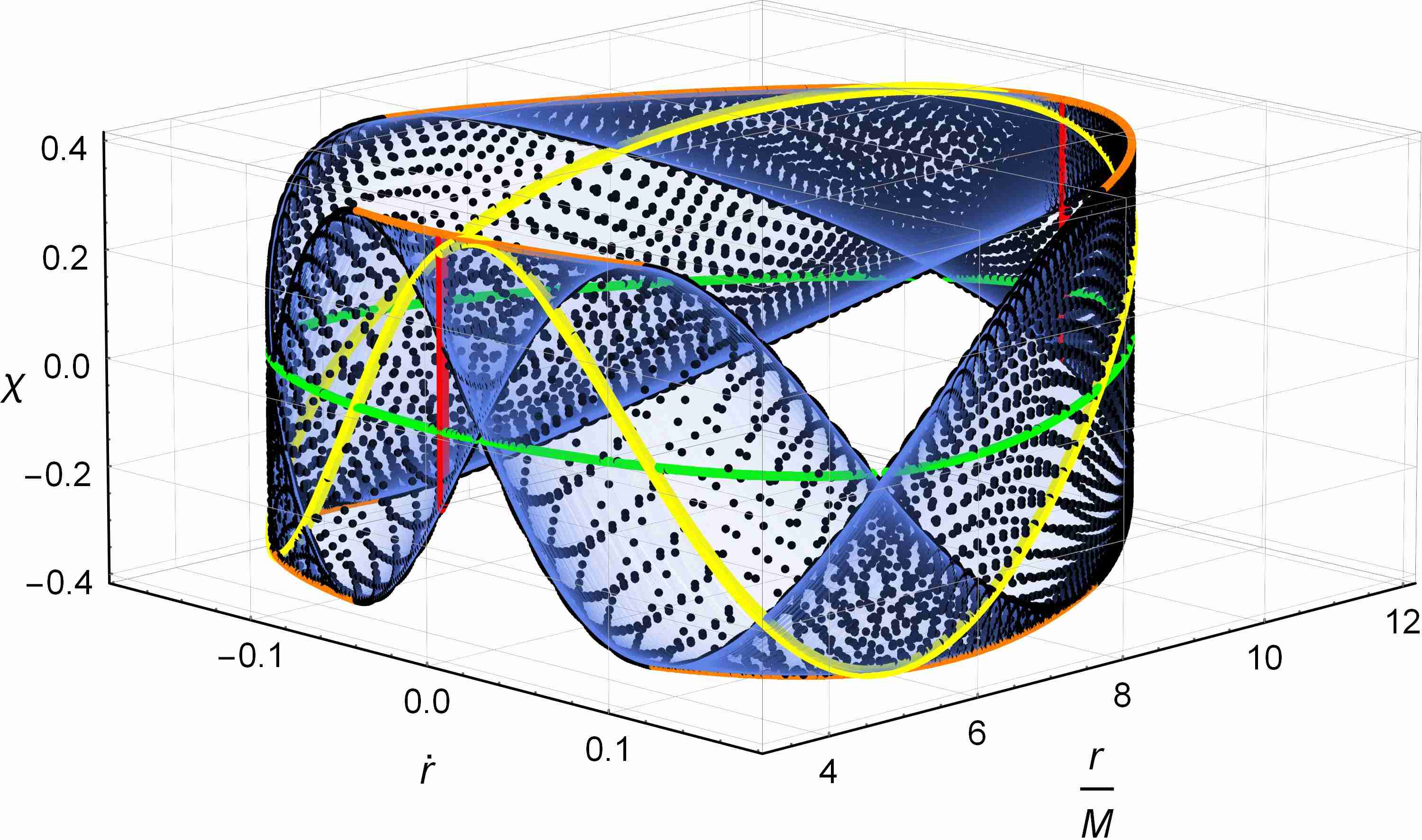}}
    \subfigure[Metric of Eq.~\eqref{metric} with $l_\textrm{NP} = 0.4M$ and $\zeta = 1$]{\includegraphics[scale=0.084]{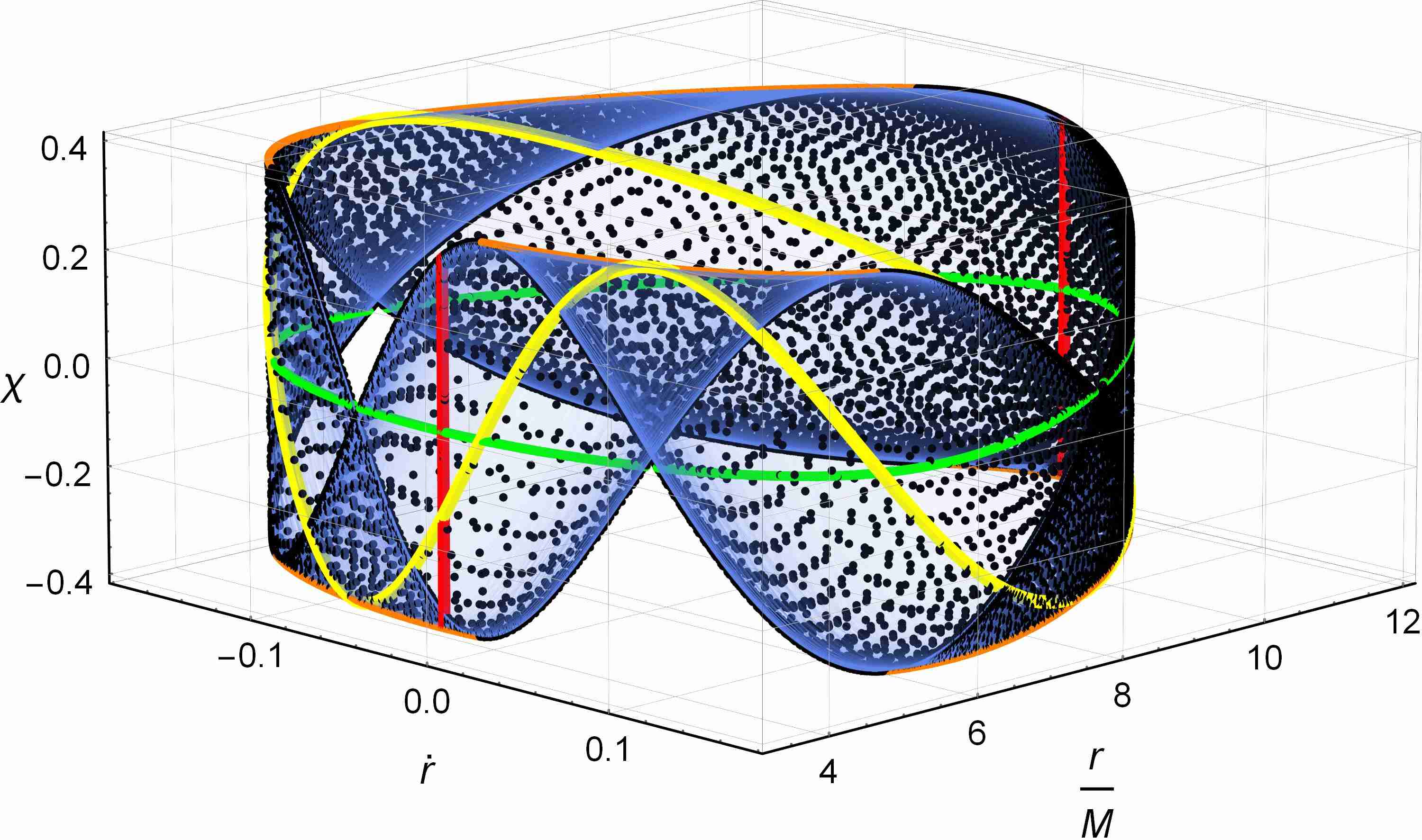}}  \\
    \subfigure[DSK metric with $\alpha_Q = 0.01$]{\includegraphics[scale=0.084]{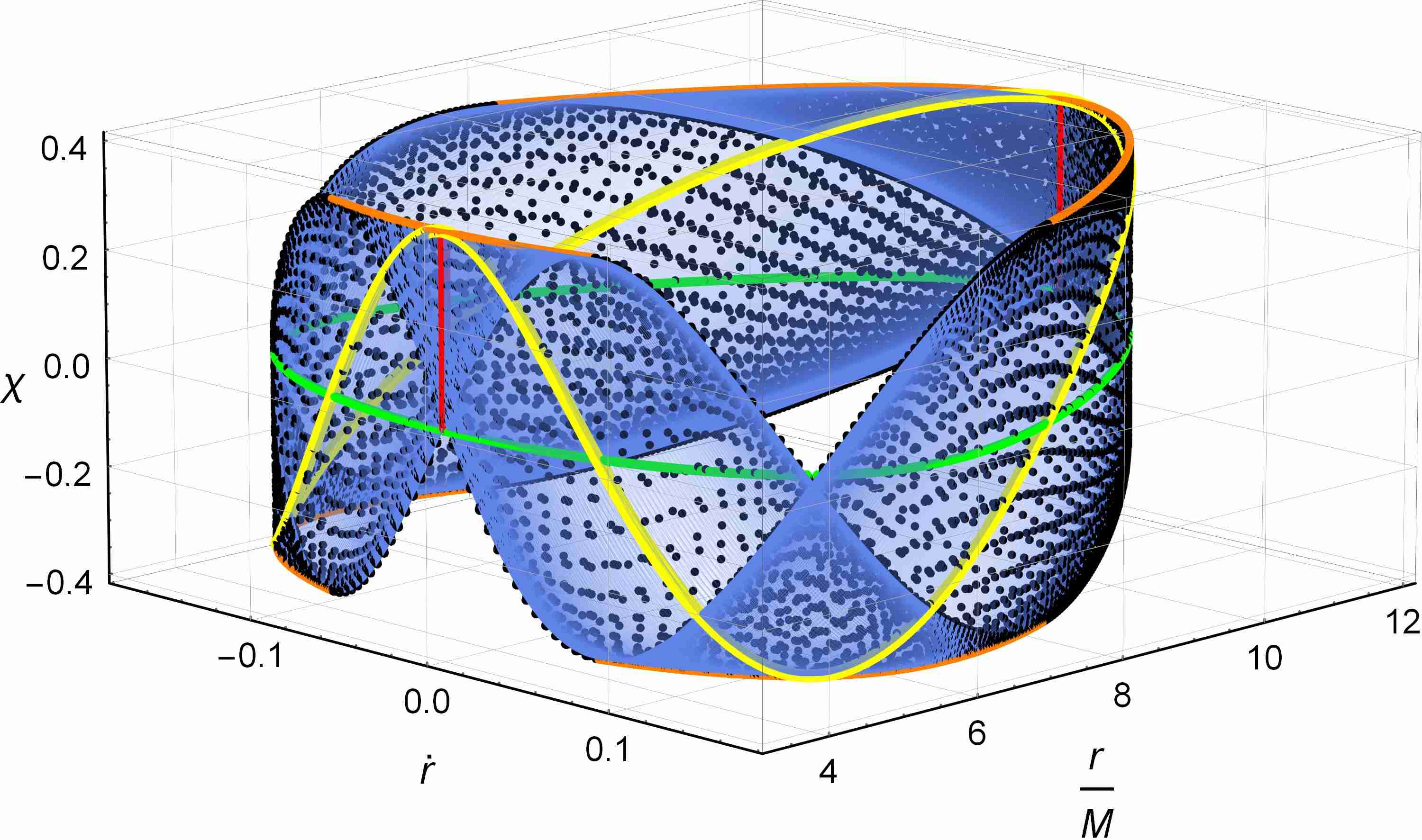}}
    \subfigure[DSK metric with $\alpha_Q = -0.01$]{\includegraphics[scale=0.084]{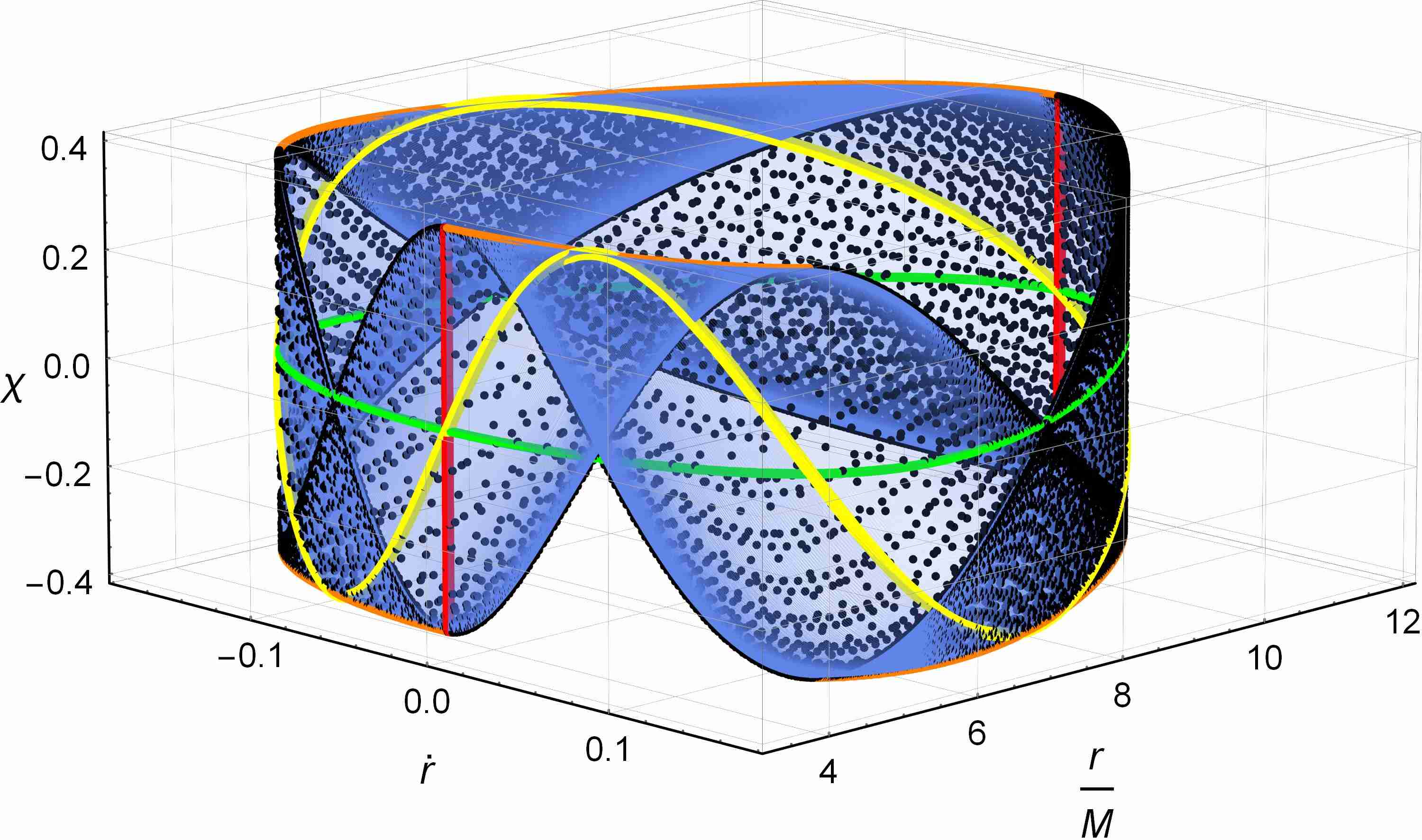}}
    \caption{Phase space trajectory of geodesics around the island, with each subfigure corresponding to different spacetimes detailed in the subcaption. The geodesics are chosen to be just inside the island (blue, with black dots equally spaced along geodesics) and at the center of the island (yellow). The colored points are the Poincar\'e sections of the geodesics with section of $\dot r = 0$ (red), section of $\chi = 0$ (green), and section of $\dot\chi = 0$ (orange), respectively.}
    \label{fig:PS3D}
\end{figure*}

In order to identify the orbit that corresponds to the center of islands in a perturbed system, we utilize the fact that the total change to the Carter constant must vanish after one period if the orbit is closed in phase space, i.e., resonating. Such an orbit can be identified by varying $\chi(0)$. In the case of the DSK metric, the orbit can be identified straightforwardly by its symmetry. Since the $(\chi,\dot r)$ inversion leaves the change rate of the Carter ``constant'' intact, the contribution from the $\dot r >0$ branch and the $\dot r <0$ branch must cancel each other. Therefore, either $\dot\chi(0) = 0$ or $\chi(0) = 0$ has to be satisfied. Furthermore, since the larger the Carter constant gets, the quicker $r$ revolves relative to $\chi$ \footnote{If we treat the right-hand side of Eqs.~\eqref{eq:r_cycle} and \eqref{eq:chi_cycle} as an effective kinetic energy, $C$ contributes positively to the total energy in Eq.~\eqref{eq:chi_cycle}, increasing $\chi$ period, and negatively in Eq.~\eqref{eq:r_cycle}.}, the fixed point is stable if $\dot C$ grows when $r$ lags behind $\chi$ cycle and vice versa. This only happens when $\alpha_Q \chi \dot\chi$ transits from positive to negative around the periapsis. Therefore, the stable fixed point at $\dot r = 0$ must reach the latitudinal turning points if $\alpha_Q > 0$ and the equatorial plane if $\alpha_Q < 0$. We demonstrate the branch-selecting rule by showing the phase space trajectories of geodesics inside the $1/2$-resonant islands of various metric setups in Fig.~\ref{fig:PS3D}, including the DSK metric with opposite signs of $\alpha_Q$.

In the case of non-Kerr spacetimes of Eq.~\eqref{metric}, however, the situation is far more complicated. As demonstrated in the top panel of Fig.~\ref{fig:DeltaC}, the asymmetry shifts the peak of $\dot C$ horizontally by $\dot r \sim 0.01$. One may be tempted to alter $\chi(0)$  equivalently as a shift in the radial cycle that matches the said amount to balance $\dot C$, and it would be done so if all four branches of the geodesic labeled by the sign of $\chi$ and $\dot\chi$ are shifted with equal contribution. Unfortunately, due to the symmetry presented at the Hamiltonian level, the edge of the invariant tori, even without circularity, shown as the allowed region of the geodesic by varying the initial condition, is still symmetric with respect to $\dot r$ inversion. The branch-crossings thus shift unevenly, leading to uneven contributions to the variation of the Carter constant among the four branches of $\chi$, $\dot\chi$ signatures. Due to the radial dependence of $M_{\zeta \to 0}$, the two secondary branches contribute about $1/2$ of the main branches. Therefore to cancel out the asymmetry, one has to double the shift to $\dot r \sim 0.02$ at $\dot\chi = 0$ or $\chi = 0$, which is half of the critical velocity of the $\zeta>0$ case, as shown in Fig.~\ref{fig:hpointzeta}. We may derive the corresponding location of periapsis, which turns out to be at $ \chi = 0.322...$ for the case of $\dot\chi = 0$, extremely close to $\chi_s$ of $\zeta<0$, as shown in Fig.~\ref{fig:R_Vs_zeta}.

Notice that the relation between the stable fixed point and the choice of the branch is opposite of what we identified in the DSK case. We cannot explain why this is happening, except that numerical integration does suggest so.

\section{Implications}\label{sec.implication}

Let us briefly discuss the possible implications of what we have found. For noncircular spacetimes, as shown in Figs.~\ref{fig:hpointzeta} and \ref{fig:R_Vs_zeta}, across $\zeta=0$ where the spacetime has integrable orbital dynamics, expectedly, the width of islands $R$ shrinks to zero and the value of $r_c$ varies smoothly. However, the amount of the asymmetry of islands changes abruptly. This means that when an integrable Hamiltonian receives nonintegrable perturbations, resonant points split and form resonant islands in a way sensitive to the type of perturbations. Specifically, unlike nonintegrable but circular spacetimes in which islands respect the discrete symmetries of the system, noncircular ones break the symmetries, manifesting as abrupt deformations of islands. As is well known, the whole bunch of islands for a particular resonance encases a stable periodic orbit which appears as the central point of islands, i.e., the stable fixed point, on the surface of section. This stable periodic orbit always appears at the center of the nested curves inside islands, irrespective of the choice of $\chi$ for a Poincar\'e section. For example, the centers of the islands in both Figs.~\ref{fig:islandsandh} and \ref{fig:SymmetricIsland} correspond to the orbit defined by the same initial condition $(r(0)=3.34888, \chi(0)=-0.15064, \dot{r}(0)=0)$. The central periodic orbit is a stable fixed point on the Poincar\'e section, and, therefore, the abrupt shift of the islands may originate from the shift of the central periodic orbit.

Given the symmetry of resonant orbits inside the islands on the section at the latitude of symmetry $\chi_s$, the radial turning points of the central periodic orbit have to reside on $\chi=\chi_s$ as well. The main distinction between circular and noncircular spacetimes is that symmetric islands appear either at $\chi_s=0$ or at the latitudinal turning points for the circular case, but $\chi_s$ can be anything for the noncircular case. Thus, in the case of noncircular spacetimes, the radial turning points of the central periodic orbit seem to acquire an off-equatorial shift, and the amount depends on the parameters of the metric. For a particular parameter choice of the metric, i.e., $a=0.66M$, $\zeta=1$, and $l_\textrm{NP}=0.4M$, these radial turning points of the central periodic orbit shift toward the negative $\chi$. As a consequence, the central periodic orbits in noncircular spacetimes would acquire asymmetric temporal expanse with respect to the equatorial plane.

The asymmetric islands, or explicitly, the off-equatorial shift to the central periodic orbits, could be interpreted as due to an off-equatorial net force acting on the orbits. This net force is a feature of noncircularity. Consider an EMRI system with a small spinning compact object gradually spiraling into a massive black hole with noncircular spacetime, and suppose that the spin of the smaller object is initially (anti-)aligned to the orbital angular momentum. The off-equatorial net force would cause the parallel spin component to partially get converted into a perpendicular component so that a nonzero spin component orthogonal to orbital angular momentum emerges \cite{Datta:2020axm,Schmidt:2014iyl}. Therefore, the evolution of the parallel component of the secondary's spin in the noncircular background should deviate from that in the Kerr background, though the detectability of such deviation through LISA is still not clear \cite{Piovano:2021iwv,Skoupy:2023lih}. In any case, if the off-equatorial net force can be quantified independently through feasible observable, one may further infer the position of the $1/2$-resonant islands and even that of the stable fixed points. This could be achieved through the measurements of the time the small object of EMRIs spends inside resonances, with sufficient events and statistics.

The way EMRI systems may help to investigate resonant orbits is the following. Typical EMRI models utilize the adiabatic approximation in which the constants of motion are assumed to evolve very slowly along the orbital motion \cite{Barack:2018yvs,Hinderer:2008dm}. In this approximation, the trajectory of the small object follows the geodesic for several periodic cycles up to a timescale at which the radiation reaction starts picking up. After one such time, the constants of motion should be updated by self-force effects averaged over $(r,\theta)$ torus. The self-force term has a phase like $k_r q^r+k_{\theta} q^{\theta}+k_{\varphi} q^{\varphi}$, where $q^A$'s are angle variables conjugate to the conserved quantities $(E,L_z,C,\mu)$ in the action-angle formalism. In non-resonant regimes, the phase oscillates rapidly and vanishes after taking averages over a radiation reaction time. On the contrary, during the resonance, since $k_r\omega^r+k_{\theta}\omega^{\theta}$ can be zero for some combinations of $k_r$ and $k_{\theta}$, the phase could evolve slowly and be non-vanishing after averaging, causing cumulative dephasing effects (around the order of $\sqrt{M/\mu}$) in the gravitational waveforms \cite{Speri:2021psr}. If these values can be extracted from multi-event analysis that calibrates and aggregates different phases inside short resonance windows of each event, one may understand how $q^r$ and $q^{\theta}$ are spread over the $(r,\theta)$ torus and further distinguish noncircular resonances from circular ones.

In addition to the asymmetric islands and the existence of off-equatorial net forces, we would like to emphasize that, for noncircular spacetimes, the amount of asymmetry of islands, i.e., $\textrm{v}_c$ and $\chi_s$, appears discontinuous at the integrable limit. The discontinuous jump of $\textrm{v}_c$ and $\chi_s$ at the integrable limit could have a crucial observational implication. Suppose a Kerr black hole receives minuscule noncircular deformations. Although the Birkhoff islands shall be narrow, $\textrm{v}_c$ and $\chi_s$ may acquire sizable values abruptly. If such a sizable value is directly associated with some observables and can be measured, perhaps through the methods mentioned above, it could be a powerful tool to test the circularity of the black hole spacetime.      

Presently, it is still not clear what the associated observables would be. Naively, such observables should, at least statistically, be able to infer the location of resonant islands. Suppose the observables can be parametrized as
\begin{equation}
\mathcal{O}=\mathcal{O}(R,\textrm{v}_c,\chi_s,X)\,,
\end{equation}
where $X$ represents other dependencies unrelated to the supermassive black hole spacetime geometry. Note that $\mathcal{O}$ has an implicit dependence on $M$, $a$, $l_\textrm{NP}$, and $\zeta$ through $R$, $\textrm{v}_c$, and $\chi_s$. If an observable depends on the island width $R$ in a way that it smoothly approaches zero toward the integrable limit, then it becomes challenging to test circularity through the measurement of said observable, even when $\textrm{v}_c$ and $\chi_s$ have sizable values. On the other hand, if an observable acquires sizable nonzero values directly determined by $\textrm{v}_c$ and $\chi_s$ near the integrable limit, it can be a perfect tool to test the circularity of black hole spacetimes. It requires further exploration to understand whether the observable $\mathcal{O}$ belongs to the former case, which could lead to a no-go for testing circularity through the asymmetry of islands, or it belongs to the latter case. It has been shown in Refs.~\cite{Destounis:2021mqv,Destounis:2021rko,Destounis:2023gpw} that the existence of chaotic orbital dynamics may leave imprints such as glitches on the spectrogram of the gravitational waves emitted by EMRIs. Variations of such imprints may serve as the observable to capture the asymmetry of the islands.



\section{Conclusions}\label{sec.conclusion}

The future space-based gravitational wave detectors, such as LISA, are going to create a new arena as precision tests of the Kerr hypothesis become possible. In particular, by examining the gravitational waves associated with the orbital dynamics of EMRI systems, one could possibly tell whether the orbital dynamics around the supermassive black hole is chaotic \cite{Destounis:2021rko,Destounis:2021mqv,Apostolatos:2009vu,Lukes-Gerakopoulos:2010ipp,Destounis:2023khj,Destounis:2023gpw}. If any signs of chaos are detected, they would strongly hint toward physics beyond GR, as the orbital dynamics is integrable and non-chaotic in the Kerr spacetime preferred by GR. 

In this paper, we investigate the chaotic features of orbital dynamics induced by black hole spacetimes that break circularity. From the analytic point of view, noncircularity prohibits any proper definition of two-surfaces that are everywhere orthogonal to the surfaces of transitivity, i.e., the surfaces spanned by the Killing vectors. Two surfaces entangle in a nontrivial way and destroy the Liouville integrability of geodesics. Numerically, upon choosing a specific noncircular spacetime of Eq.~\eqref{metric}, we exhibit the existence of chaos by identifying the resonant islands in the Poincar\'e surfaces of section. A similar analysis has been done in Ref.~\cite{Zhou:2021cef} in which the authors considered a different noncircular spacetime and confirmed the existence of chaos in orbital dynamics.

In addition to confirming the existence of chaos, in this paper, we find that noncircularity not only gives rise to chaos but also induces chaotic features unheard of in typical nonintegrable but circular spacetimes. More explicitly, the resonant islands appear asymmetric with respect to the $\dot{r}=0$ axis on the $(r,\dot{r})$ Poincar\'e surface of section at the equatorial plane. The perturbation analysis conducted in sec.~\ref{sec.pertur} shows that such asymmetric patterns also happen to non-resonant orbits. The orbits in the phase space lose the $(\dot{r},\chi)$ symmetry precisely due to the surface of transitivity not properly foliating the noncircular spacetime. 

To quantify the asymmetry of resonant islands, we define two measures in sec.~\ref{sec.numerical}, the critical velocity $\textrm{v}_c$ and the latitude of symmetry $\chi_s$. These two measures are uniquely defined for a given set of $E$, $L_z$, and the metric parameters. As opposed to circular but nonintegrable spacetimes, these two measures are in general not zero for noncircular spacetimes. Interestingly, we find that $\textrm{v}_c$ and $\chi_s$ acquire discontinuous jump at the parameter space where the geodesics are integrable (see Figs.~\ref{fig:hpointzeta} and \ref{fig:R_Vs_zeta}). It implies that the asymmetry of islands, or more explicitly, the way that central periodic orbits develop islands, is sensitive to the form of noncircularity rather than to its strength. This is in contrast to the size of islands that depends heavily on the strength of chaos because the islands shrink toward the integrable limit following the KAM theorem (see the behavior of the island width $R$ in Fig.~\ref{fig:R_Vs_zeta}).  

We would like to emphasize that although we choose the metric in Eq.~\eqref{metric} as an example to demonstrate our results, we expect our conclusions to be genuine in the presence of noncircularity. One timely extension of this work is to identify physical observables that link to the asymmetric patterns of islands. In sec.~\ref{sec.implication}, we lay down some preliminary statements regarding possible observational implications of noncircularity. It becomes timely to investigate the possibility of inferring the asymmetric distribution of islands from gravitational wave data or from possible but yet-to-know features of noncircularity in the gravitational wave glitches \cite{Destounis:2021rko,Destounis:2021mqv,Destounis:2023khj,Destounis:2023gpw}. Due to the discontinuous jump at the integrable limit, an extremely tiny amount of noncircularity would already give rise to sizable effects of this kind. Therefore, if we may infer the position of islands inside the phase space from gravitational wave data, it would be a perfect tool to test the circularity of black hole spacetimes. In addition, the resonant orbits can be investigated perturbatively using the effective resonant Hamiltonian approach \cite{Pan:2023wau}. It could be possible to apply the approach, perhaps with some generalizations, to noncircular spacetimes. We leave these topics to future work.

\acknowledgments
CYC is supported by the Institute of Physics of Academia Sinica and the Special Postdoctoral Researcher (SPDR) Program at RIKEN. HWC is supported by Ministry of Science and Technology (MoST) of Taiwan grant number 110-2811-M-002-686 through department of physics and the Leung Center for Cosmology and Particle Astrophysics (LeCosPA) of National Taiwan University. AP is supported by MoST grant No. 110-2811-M-003-507-MY2 through department of physics and the Center of Astronomy and Gravitation (CAG) of National Taiwan Normal University.

\end{document}